\RequirePackage{fix-cm}
\documentclass[onecollarge,useAMS]{svjour2}
\smartqed  
\usepackage{graphicx}
\usepackage{subfig}
\usepackage{graphicx,color}	
\usepackage{verbatim}
\usepackage{array,amsmath}
\usepackage[latin1]{inputenc}
\usepackage{epsfig}
\usepackage[normalem]{ulem}
\begin{document}

\title{Dynamics of the 3/1 planetary mean-motion resonance.
}
\subtitle{An application to the HD60532 b-c planetary system}

\author{A. J. Alves \and
        T. A. Michtchenko \and M. Tadeu dos Santos 
}


\institute{A.J. Alves \and T. A. Michtchenko \and M. Tadeu dos Santos\at
          Instituto de Astronomia, Geof\'{\i}sica e Ci\^{e}ncias Atmosf\'{e}ricas (IAG) \\
	      Universidade de S\~ao Paulo, S\~ao Paulo, Brazil	
              Tel.: +55-11-30912739\\
              \email{alan.ajac@iag.usp.br}
	    }
\date{Accepted for publication at Celestial Mechanics and Dynamical Astronomy.\\ \newline The final publication is available at Springer via http://dx.doi.org/[10.1007/s10569-015-9664-x]}

\maketitle

\begin{abstract}

In this paper, we use a semi-analytical approach to analyze the global structure of the phase space of the planar planetary 3/1 mean-motion resonance, in cases where the outer planet is more massive than its inner companion. We show that the resonant dynamics can be described using only two fundamental parameters, the total angular momentum and the scaling parameter.  The topology of the Hamiltonian function describing the resonant behaviour is studied on the representative planes that allows us to investigate a large domain of the phase space of the three-body problem without time-expensive numerical integrations of the equations of motion, and without any restriction on the magnitude of the planetary eccentricities. The families of the well known Apsidal Corotation Resonances (ACR) parameterized by the planetary mass ratio are obtained and their stability is analyzed. The main dynamical features in the domains around  ACR are also investigated in detail by means of spectral analysis techniques, which allow
us to detect the regions of different regimes of motion of resonant systems. The construction of dynamical maps for various values of the total angular momentum shows the evolution of domains of stable motion with the eccentricities, identifying possible configurations suitable for exoplanetary systems.
\keywords{exoplanetary systems \and resonant dynamics \and periodic orbits}
\end{abstract}

\section{Introduction}
\label{intro}
The resonant configurations are frequent in the extra-solar planetary systems. In the studies on the formation and long-term stability of the resonant systems, a deep understanding of the behaviour of the planets becomes important. Indeed,  long-term stable motion of close planets at high eccentricities is possible only if the planets are locked in (and protected by) a mean-motion resonance (hereafter just a resonance, or MMR). Second, resonance trapping appears to be a natural outcome of planetary migration processes due to planet-disk interactions, which are believed to take place in the latest stage of the planet formation (e.g. Kley 2000, Snellgrove et al. 2001, Kley et al. 2005).

This paper is a part of the series of papers devoted to the study of the resonant dynamics of the planets. In the previous works, we have addressed 3/2 (Callegari et al. 2006), 2/1 (Callegari et al. 2004, Michtchenko et al. 2008 a,b) and 5/2 MMRs (Michtchenko \& Ferraz-Mello 2001a).  This paper focus on the 3/1 resonance planetary configuration. Although this resonance is not densely populated as those mentioned above (Antoniadou \& Voyatzis 2014) (only the HD60532 b-c system is confirmed to evolve inside it), there are still many candidates from the Kepler database waiting for confirmation (\textit{http://exoplanets.org}; Han et al. 2014).

The  resonant dynamics may be partially understood through the study of the special solutions often referred to as Apsidal Corotation Resonance (ACR). Several previous papers were devoted to the determination of these periodic solutions in the planar 3/1 MMR  (Beaug\'e et al. 2003, Ferraz-Mello et al. 2003, Michtchenko et al. 2006,  Voyatzis \& Hadjidemetriou 2006). Voyatzis (2008) complemented the study of the ACR solutions analyzing their dynamical stability and constructing the dynamical maps for the case of the 55Cnc planets.  In more recent paper (Antoniadou \& Voyatzis 2014), the study of the families of ACRs was extended to the nonplanar dynamics of the planets.

Notwithstanding the attention devoted to ACRs and periodic orbits, not much is known on the topology of the phase space of the 3/1 MMR. Moreover, since most of the detected exoplanets evolving in resonances are outside the immediate vicinity of the ACRs (this is the case of the HD60532 planets \textbf{b} and \textbf{c}), the dynamical exploration of the whole phase space of the resonant problem becomes important, from the point of view of stability of the planetary motion.  That is the purpose of the present paper.
\begin{figure}
\centering
\includegraphics[width=0.8\textwidth]{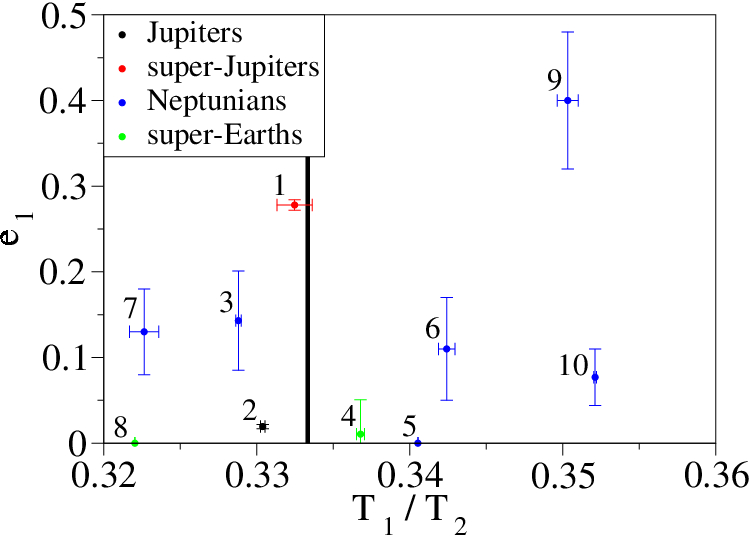}
\caption{The distribution of the extra-solar planetary systems near the nominal 3/1 MMR. The planetary systems are numbered as: 1 - HD60532\,b-c; 2 - 55Cnc\,b-c; 3 - HD10180\,d-e; 4 - GJ\,163\,b-c; 5 - Kepler 20\,b-c; 6 - HD20781\,b-c; 7 - HD31527\,b-c; 8 - Kepler 11\,b-e; 9 - HD20003\,b-c; 10 - HD10180\,c-d. Different colors are chosen to identify the averaged planetary masses of the systems (see text for details).}\label{fig:planetary_systems31}
\end{figure}

The paper is organized as follows. In the next section, we do a brief overview of the exoplanetary systems detected near the 3/1 commensurability. The hypothesis on the origin of this configuration through the capture of planetary pairs during migration stages is also discussed in this section. In Section \ref{themodel}, we introduce the basic semi-analytical Hamiltonian model which describes the planar 3/1 resonant problem. The topology of the resonant Hamiltonian is studied in Section \ref{sec-3}, where we plot energy levels on the representative planes, calculate the symmetric and asymmetric ACRs of the 3/1 MMR and discuss the Law of Structure. The dynamics around the ACRs is investigated in Section \ref{dyn_out_acr}, in form of dynamical maps and dynamical power spectra, and the main regimes of motion of the planets in the 3/1 MMR are described. Finally, the conclusions are done in Section \ref{conclus}.

\section{Exoplanets near the 3/1 MMR}\label{sec-1}

Analyzing the orbital configurations of the planetary systems from the Exoplanet Orbit Database (Han et al. 2014), we have identified 10 pairs with ratio of orbital periods from the interval $2.85<T_2/T_1<3.15$ ($T_1$ and $T_2$ are the orbital periods of the inner and outer planets, respectively). They are shown in Figure \ref{fig:planetary_systems31} by solid circles on the ($T_1/T_2$,$e_1$)--plane, where $e_1$ is the eccentricity of the inner planet's orbit. The nominal location of the 3/1 MMR on this plane is shown by the vertical line. The systems are classified according their averaged masses defined as $M=0.5(m_1+m_2)$, where $m_1$ and $m_2$ are the masses of the inner and outer planets, respectively. The systems with averaged masses from the range between two and ten times the mass of the Earth are referred to as \text{super-Earth} and are shown by green symbols, the systems from the range $0.02 M_\textrm{J} < M < 0.2 M_\textrm{J}$ ($M_\textrm{J}$ is the mass of Jupiter) as \text{
Neptunians} (blue symbols), from $0.2 M_\textrm{J} < M < 2 M_\textrm{J}$ as \text{Jupiters} (black symbols) and those with the masses $M > 2 M_\textrm{J}$ as \text{super-Jupiters} (red symbols).

The closest to the nominal 3/1 MMR is the HD60532  system (\# 1 in Figure \ref{fig:planetary_systems31}), which consists of the F-type star of mass $M_{\star}=1.44 M_\odot$ and two planets, whose co-planar orbits were first determined through a Keplerian fitting of 147 radial velocity spectra in Desort et al. (2008). The dynamical analysis of the system pointed toward a possible 3/1 mean-motion resonance, however, the weak, but significant, instabilities were also detected in the two-planet motion. Laskar \& Correia (2009) applied a Newtonian N-body fit, still assuming co-planar motion, but including the inclination with respect to the plane of sky as a free parameter. They obtained the slightly improved best-fit configuration with the inclination of $20$ degrees.  With this inclination, the planets become super-Jupiters, with masses $m_1=3.1548 M_{J}$ and $m_2=7.4634 M_{J}$, for inner and outer bodies, respectively, and the system's 3/1 MMR evolution is stable, at least, over 5 Gyr.

The physical and orbital parameters of the HD60532 system obtained in Laskar \& Correia (2009) are shown in Table \ref{Table:hd60532bc_data} and will be used in applications of the developed model in this paper. However, some preliminar considerations concerning their reliability must be done. Indeed, the confidence intervals listed in Table \ref{Table:hd60532bc_data} seem to be underestimated. In the orbit determination, the error bars are usually calculated through the Monte Carlo techniques, for instance, \textit{Biased-Monte-Carlo Method (BMC)}, as described in Tadeu dos Santos (2012). Beaug\'e et al. (2012) applied the BMC method to determine the orbits of HD60532 and estimated the possible ranges of the eccentricities, supported by the set of observations, as [0.2 - 0.35] and [0.0 - 0.1], for planets b and c, respectively (see Figure 3 in that paper). For the orbital periods, these ranges are [200-205] and [570-640] days, respectively.  Within the confidence intervals given in Beaug\'e et al. (
2012), the system can be found in qualitatively distinct configurations, either resonant or near-resonant ones. On the other hand, the narrow confidence intervals reported in Laskar \& Correia (2009) seem to be determined by the dynamical criterion (see Figure 2 in that paper), that constrains the system to evolve inside the 3/1 MMR.

The dynamical stability of the system will depend also on the assumed masses of the planets: for example, S\'andor \& Kley (2010) claim that the simulations of the formation of the HD60532 planets suggest that the capture into the 3/1 MMR takes place only for higher planetary masses, thus favouring orbital solutions having small inclinations ($I \sim 20^\circ$),  found in Laskar \& Correia (2009). This fact can be easily understood from the theories of the resonant dynamics which guarantees the larger width of the resonance for higher planet masses and protects the eccentric planets from close approaches and collisions (Michtchenko et al., 2008\,a,b).

Thus, we still can use the orbital configuration of the HD60532 planets shown in Table \ref{Table:hd60532bc_data} as the case study to illustrate the dynamics of a planetary system inside or near the 3/1 MMR, which is stable over $5$G years. Reminding that the focus of this paper is the dynamical analysis of a planetary system inside the 3/1 MMR, the option for Laskar \& Correia (2009) solution is intentional to fulfill our purpose.

\begin{table}
 \centering
 \begin{tabular*}{0.55\textwidth}{c c c c}
 \hline
 \hline
\textbf{Param.} & \textbf{[units]} &\textbf{HD60532\,b} & \textbf{HD60532\,c} \\ \hline
\multicolumn{1}{c}{$\mathit{\sqrt{\chi^2}}$} & & \multicolumn{2}{c}{$4.369$} \\
rms      & [m/s] & \multicolumn{2}{c}{$4.342$} \\ \hline
Date  &	[JD-$2400000$]& \multicolumn{2}{c}{$54000.000$(fixed)} \\	
$V$  &	[km s$^{-1}$] & \multicolumn{2}{c}{$-0.0055\pm 0.0003$} \\
$P$ &	[day] & $201.83\pm 0.14$ & $607.06\pm 2.07$\\
$\lambda$ & [deg] & $14.78\pm 0.66$ & $317.02\pm 0.93$\\	
$e$ &	& $0.278\pm 0.006$ & $0.038\pm 0.008$\\
$\omega$ & [deg] & $352.83\pm 1.05$ & $119.49\pm9.14$\\ 	
$K$ & [m/s] & $30.34\pm0.32$ & $47.84\pm 0.44$\\
$I$ & [deg] & $20$(fixed) & $20$(fixed)\\ \hline
$M$ & [$M_{jup}$] & $3.1548$ & $7.4634$ \\
$a$ & [AU] & 0.7606 & 1.5854 \\ \hline
\end{tabular*}
  \caption[Data of HD60532 b-c]{Orbital parameters of the HD60532\,b-c system. The mass of the central star is $M_\star=1.44M_\odot$. $I$ is the inclination of the co-planar orbits in relation to the sky plane. The data are taken from Laskar \& Correia, 2009.}
  \label{Table:hd60532bc_data}
\end{table}

The planetary pairs that are close, but not inside the 3/1 MMR, are 55\,Cnc\,b-c (\# 2 in Figure \ref{fig:planetary_systems31}; Nelson et al. 2014), HD10180\,d-e (\# 3; Lovis et al. 2011), GJ\,163\,b-c (\# 4; Bonfils et al. 2013) and Kepler 20\,b-c (\# 5; Gautier et al. 2012). Initially, the pair b-c of the 55\,Cnc system was supposed to evolve inside a 3/1 MMR (Zhou et al. 2004, Marzari et al. 2005). However, since the orbital elements of the planets were modified substantially  by further observations, their dynamics was recalculated and the planetary system was found to evolve in a so-called near-resonant regime of motion (Dawson \& Fabrycky 2010).

HD10180 is a solar type star, with $M_\star=1.06M_\odot$. It hosts multi-planetary system (Mayor et al. 2011, Tuomi 2012, Kane \& Gelino 2014). Although not all components of this system are still confirmed, it would be interesting to compare the dynamics of the planets `d' and `e', evolving near the 3/1 MMR, to the near-resonant structure of our solar system, particularly, with the dynamics of the Saturn-Uranus pair, also close to the 3/1 MMR (Michtchenko \& Ferraz-Mello 2001b).

Several recently detected exoplanetary pairs from the Kepler database also present orbital periods close the 3/1 commensurability, e.g. Kepler-180 b-c, Kepler-326 b-d, Kepler-359 b-d, Kepler-107 b-e, Kepler-102 b-c, Kepler-373 b-c, Kepler-84 c-d, among others. However, the lack of information on the physical and orbital parameters of the Kepler exoplanets makes a deep analysis of their dynamics unfeasible.
\begin{figure}
  \centering
  \includegraphics[width=0.7\textwidth]{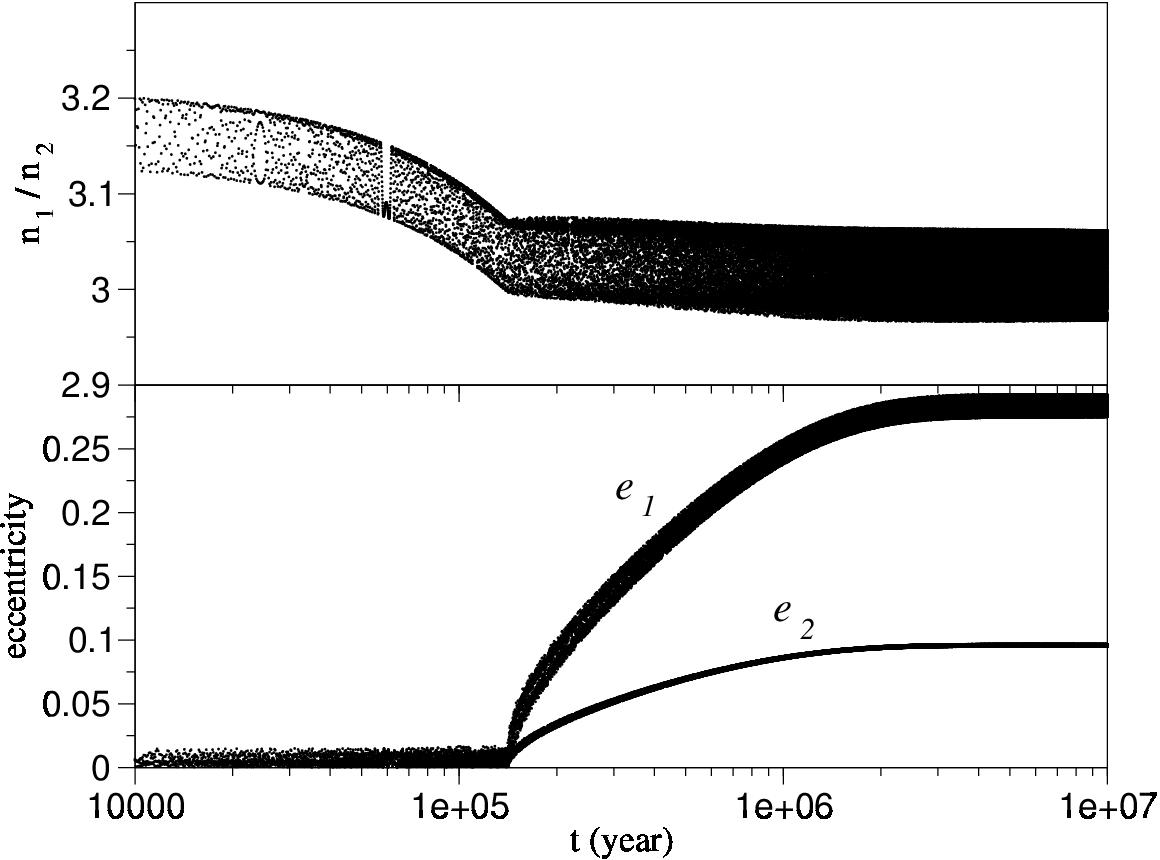}\\
  \includegraphics[width=0.7\textwidth]{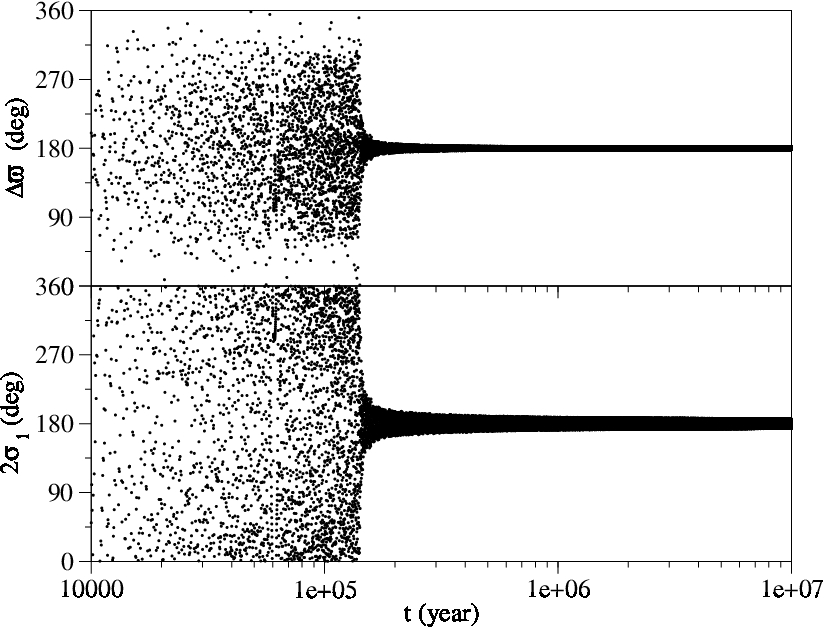}\\
  \caption{The resonance capture of the HD60532 planets \textbf{b} and \textbf{c}  evolving under the dissipative Stokes-like force (\ref{stokes_eqn}). Starting at $n_1/n_2 \sim 3.15$, the planets are ultimately trapped in the 3/1 MMR after $2\times 10^5$ years of the evolution. Top panel: Time evolution of the ratio of the mean motions (top) and the eccentricities (bottom). Bottom panel: Time evolution of the secular angle $\Delta\varpi$ (top) and the resonant angle $2\sigma_1$ (bottom), both oscillating around $180^\circ$ in the resonance.
}\label{fig:mig2_zeroecc}
\end{figure}

\subsection{On the origin of the 3/1 resonant configuration in the HD60532 system}

The main mechanism to forming a resonant planetary system is thought to be by convergent migration of planets during the final stage of their formation and consequent capture inside the resonance. Migration is the result of dissipative gravitational interactions of a protoplanetary disk and the growing planets (Lin \& Papaloizou 1979, Goldreich \& Tremaine 1979, Goldreich \& Tremaine 1980). This scenario of the origin of the  HD60532 pair has been investigated  in S\'andor \& Kley (2010), where the locally isothermal version of FARGO software was applied to simulate disk-planet interactions. However, the authors have faced some difficulties to simulate the formation of two giant planets in the 3/1 mean-motion resonance, probably, due to incorrect assumption on the final \textit{asymmetric} configuration of the planetary orbits.

To verify the possibility of the capture  of the migrating planetary pair in the 3/1 MMR, in this work, we perform a simple experiment,  modeling gravitational disk-planet interactions with a Stokes-type non-conservative force (Ferraz-Mello et al. 2003, Beauge et al. 2006, Michtchenko \& Rodr\'iguez 2011). The dissipative force component can be added to the Newton exact equations of motion of a planet as
 \begin{equation}
  \dfrac{d^2 \textbf{r}}{dt^2}= -C(\pmb{\textbf{v}}-\alpha \pmb{\textbf{v}}_c),
  \label{stokes_eqn}
 \end{equation}
where $\textbf{r}$ and $\textbf{v}$ are the position and velocity vectors of the planet,  $\textbf{v}_c$ is the circular velocity vector at the same position, while $C$ and $\alpha$ are parameters characterizing the disk in which the planet is migrating. As shown in Beaug\'e \& Ferraz-Mello (1993) and Gomes (1995a,b),  up to first order in eccentricities, the effect of the force (\ref{stokes_eqn}) on the planetary semimajor axis and eccentricity  behaviour is given by
\begin{equation}
 {a}(t)={a_{0}}\operatorname{e}^{-A t},\hspace{1cm}
 {e}(t)={e_{0}}\operatorname{e}^{-B t},
\label{stokes_exp}
 \end{equation}
where $a_{0}$, $e_{0}$ are the initial values of the semimajor axis and eccentricity, respectively, and $|A|$ and $|B|$ are inverse of the e-folding times in each orbital element, $\tau_a$ and $\tau_e$, respectively. These quantities are associated with the parameters of the Stokes force (\ref{stokes_eqn}) through:
 \begin{equation}
A=2C(1-\alpha), \hspace{1cm} B=C\alpha.
\end{equation}

Using this 'toy' model for the driving mechanism, we performed a series of numerical simulations of the resonance capture of the HD60532 planets. Several initial configurations were tested and the results have shown that the capture inside the 3/1 MMR is very robust. One of the simulations is shown in Figure \ref{fig:mig2_zeroecc}. The HD60532 pair of planets was initially placed on nearly circular orbits beyond the 3/1 MMR, at $n_1/n_2\sim 3.20$. The adopted values of the e-folding times, $\tau_a$ and $\tau_e$,  were $5\times 10^6$ and $2.5\times 10^5$ years, respectively. The dissipative force (\ref{stokes_eqn}) was applied only on the outer planet.

Figure \ref{fig:mig2_zeroecc} shows that, after the approximately $2\times 10^5$ years of the decay  of the outer planet, the pair of planets is captured in the resonance, when the ratio of the mean motions is trapped around 3/1 (top panel). The eccentricities of the planetary orbits  then increase rapidly and reach their current values soon after $2$ millions of years. The behavior of the characteristic angular variables, the secular angle $\Delta\varpi$ and the resonant angle $2\sigma_1$ (bottom panel), confirms that the system has reached the state of the minimal energy characterized by anti-aligned pericenter lines and the oscillating around $180^\circ$ critical angle of the 3/1 MMR (the definition of the angular variables is done in the next section).

\section{Modeling the 3/1 MMR}\label{themodel}

Our model consists of a system with two planets, with the masses $m_1$ and $m_2$, orbiting a central star with the mass $m_0$. The index $"1"$ will always refer to the inner planet, while the index $"2"$ to the outer one. We assume that both orbits lie in the same plane. We apply the model to the HD60532 planets b-c, whose physical and orbital parameters are shown in Table \ref{Table:hd60532bc_data}. We use the Poincar\'e canonical angle-action variables in the astrocentric reference (Laskar 1991):

\begin{eqnarray}\label{delaunay}
\lambda_i & = & \text{mean \, longitude},         \hspace{1.1cm}  L_i = {m^\prime}_i \sqrt{\mu_i a_i},\label{trans1}\nonumber \\
-\varpi_i & = & \text{longitude of pericenter},   \hspace{0.2cm} L_i - G_i = L_i \left(1-\sqrt{1-e_i^2}\right),\,\,\label{trans2}
\end{eqnarray}
where $a_i$, $e_i$ are the semimajor axes and eccentricities of the planets, respectively,  $m^{\prime}_i=\dfrac{m_i m_0}{m_i+m_0}$ are the reduced masses and $\mu_i = G(m_0+m_i)$, where $G$ is the gravitational constant.

The Hamiltonian ${\mathcal H}$ of the problem can be expressed as the sum of the Keplerian part
\begin{equation}\label{eq:ham0}
 {\mathcal H}_0=-\sum_{i=1}^{2} \dfrac{{\mu_i}^2 {m_i^{\prime}}^3}{2 L_i^2}\\
\end{equation}
and the perturbation
\begin{equation}\label{eq:ham1}
 {\mathcal H}_1=-\dfrac{G m_1 m_2 }{\Delta}+ {\mathcal T_1}.
 \end{equation}
The first term in Eq.(\ref{eq:ham1}) is the direct part of the perturbation function, $\Delta$ is the instant distance between the planets, and the second term ${\mathcal T_1}$ is the indirect part of the perturbation function. The reader is referred to Laskar \& Robutel (1995) and Ferraz-Mello et al. (2005), for further details.

\begin{figure}
   \centering
  \includegraphics[width=0.7\textwidth]{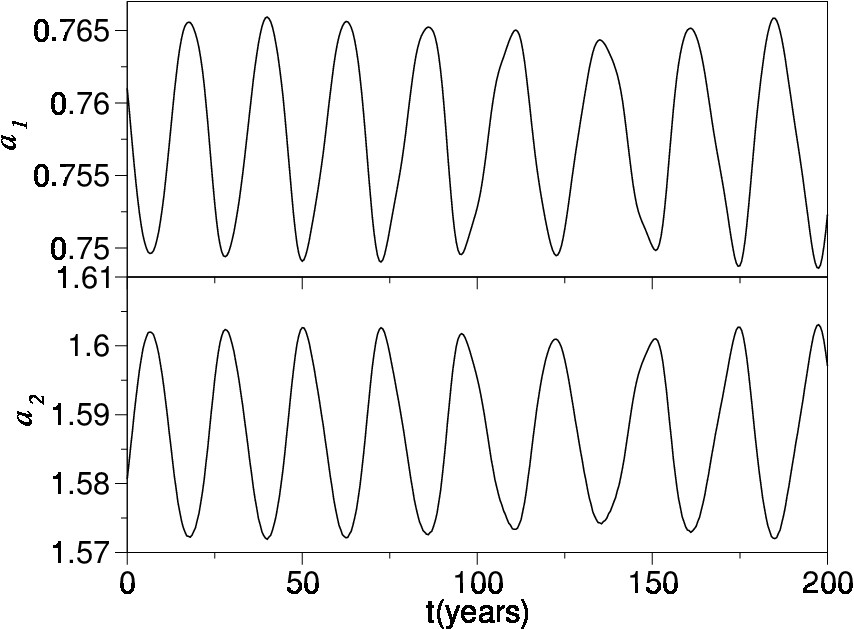}\\
  \includegraphics[width=0.7\textwidth]{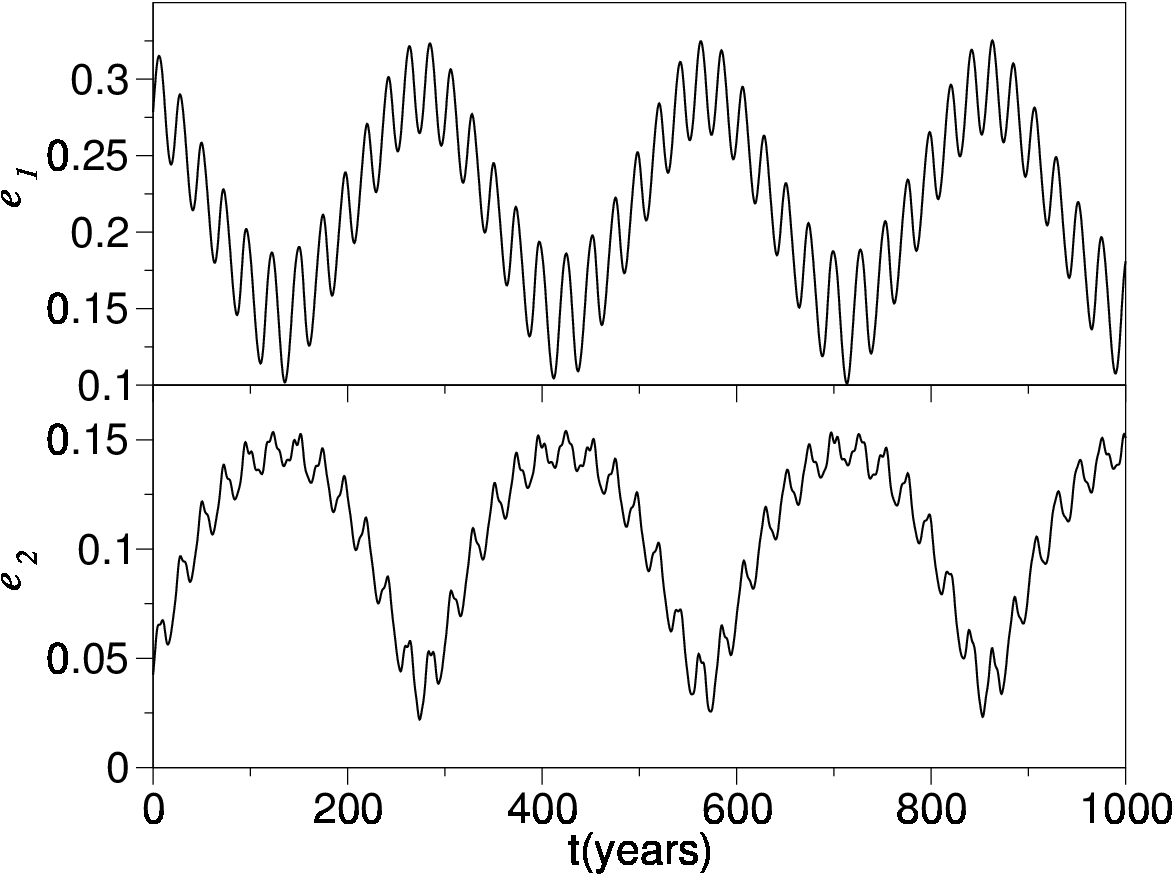}\\
  \caption{Numerical simulation of the HD60532 planets \textbf{b} and \textbf{c}, with masses and initial conditions from Table \ref{Table:hd60532bc_data}. The top panel shows the resonant oscillations of the semimajor axes, while the bottom panel shows the oscillations of the eccentricities, all averaged over short-term oscillations (of order of orbital periods). The curves were obtained by the averaged Hamiltonian equations of motion.}\label{fig:aiei_oscillation}
\end{figure}

To study the vicinity of the 3/1 MMR resonance, we transform the angular variables (\ref{delaunay}) to the critical angles of the resonance $(p+q)/p$ as
\begin{eqnarray}\label{eq:1}
\lambda_1, \qquad \qquad \qquad \qquad J_1 & = & L_1 +s\left(I_1 + I_2\right),\nonumber\\
\lambda_2, \qquad \qquad \qquad \qquad  J_2 & = & L_2 -(1+s)\left(I_1+I_2\right),\nonumber\\
\sigma_1 = (1+s)\lambda_2-s\lambda_1-\varpi_1,\qquad  I_1 & = & L_1 -G_1,\\
\sigma_2 = (1+s)\lambda_2-s\lambda_1-\varpi_2,\qquad  I_2 & = & L_2 -G_2,\nonumber\label{var1}
\end{eqnarray}
where $s=p/q$ (for the 3/1 MMR,  $p=1$ and $q=2$). In the next, the averaging of the Hamiltonian ${\mathcal H}$ is done with respect to the synodic angle $Q=\lambda_2-\lambda_1$. In the proximity of a mean-motion resonance, the variation of $Q$ is much faster than those of the resonant and secular angles, and does not influence significantly the long-term evolution of the system. Thus, all periodic terms dependent on $Q$ can be eliminated (i.e. averaged out) from the Hamiltonian function, and only secular and resonant terms need to be retained (for details, see Beaug\'e \& Michtchenko 2003).

The averaged resonant problem has two invariant quantities (or integrals of motion):  one is \emph{the spacing parameter}, written as
\begin{equation}
 {\mathcal K}=(1+s)L_1+sL_2.
\label{spacing_parameter}\\
\end{equation}
This integral describes the coupling behaviour of the planetary semimajor axes, which oscillate  with opposite phases and with amplitudes which are inversely proportional to the planetary masses. This behaviour is shown on the top graph in Figure \ref{fig:aiei_oscillation}, which shows that the timescale of the resonant oscillations is of order of tens of years. The curves were calculated through the integration of the equations of motion of the averaged Hamiltonian. Note that this behaviour of the semimajor axes is exclusively due to resonant interactions and there is no secular component in the variation of the semimajor axes (Michtchenko et al. 2011), at least, up to first order in masses.

The second integral of motion of the problem is \emph{the total angular momentum}, written as
\begin{equation}
{\mathcal AM}=J_1+J_2=L_1\sqrt{1-e_1^2}+L_2\sqrt{1-e_2^2}.
\label{amtotal}
\end{equation}
The invariance of the total angular momentum implies the coupling oscillations in the planetary eccentricity, in such a way that when the eccentricity of one planet increases, the other has to decrease. Figure \ref{fig:aiei_oscillation}\,\textit{bottom} shows the resonant evolution of the planetary eccentricities, which is composed of two components: resonant (with small amplitudes) and secular (with large amplitudes) ones. The timescale of the secular variation associated to the variation of the secular angle $\Delta\varpi=\varpi_2-\varpi_1$ is of order of hundreds of years. Note that, despite the very large planetary masses, it is still one order longer than the timescale of the resonant oscillations.
\begin{figure}
  \centering
  \includegraphics[width=0.8\textwidth]{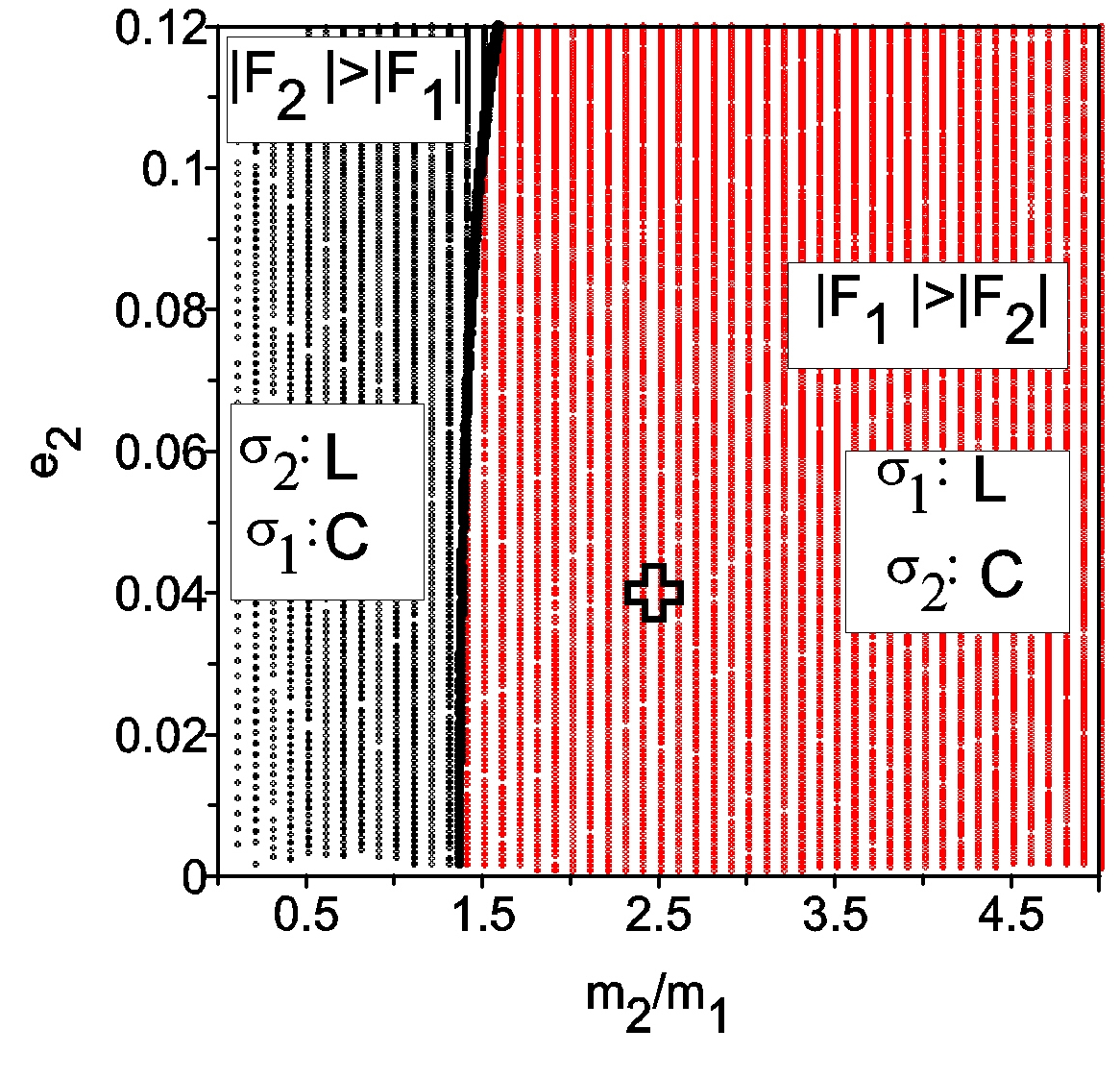}\\
  \caption{Parametric plane ($m_2/m_1$,$e_2$) of the planetary mass ratio and the outer planet's eccentricity. The black color is used to indicate the domain where $|F_2|>|F_1|$ and $\sigma_2$ is a truly resonant angle, while the red color shows the domain where $|F_1|>|F_2|$ and $\sigma_1$ is a truly resonant angle. $|F_1|$ and $|F_2|$ are amplitudes of the resonant terms in the analytical first-order expansion of the Hamiltonian (\ref{hamiltoniano_laplace31}). Cross symbol shows the location of the HD60532 system.}\label{fig:f1f2f3_laplace}
\end{figure}

\subsection{The choice of the truly resonant angle}

The invariance of ${\mathcal AM}$ and ${\mathcal K}$ has important consequences for the orbital evolution of the system. It indicates that, after the averaging process, from the set of the variables ($L_1,L_2,G_1,G_2$) (or,   $a_1,a_2,e_1,e_2$), only two are independent, and the planar resonant problem has two degrees of freedom, with two proper modes of motion. One mode is defined as \textit{a secular mode} of motion of the system and is associated with the secular angle $\Delta \varpi=\varpi_2-\varpi_1$. The typical behaviour of this angle inside the resonance domain is a circulation/oscillation (Michtchenko et al. 2011).

The other mode is \textit{a resonant mode} defined by the libration of the resonant angle. By the definition of the critical angles (Eq. (\ref{eq:1})), $\Delta\varpi = \sigma_1-\sigma_2$, that means that only one of two critical angles is independent and is a \emph{truly} resonant angular variable of the problem (the other independent angular variable is $\Delta\varpi$). The question which rises now is which one of two critical angles can be chosen as a resonant variable.

The choice can be defined invoking the limit-case of the resonant restricted three-body  problem. Indeed, the theories of the dynamics of the asteroids from the Main Belt and of the trans-Neptunian objects provide the opposite resonant angles. In the former case, when the mass of the inner body is much smaller than the mass of the outer planet, the resonant angle is $\sigma_1$.  In the last case, the librating angle is related to the outer body, $\sigma_2$. We refer to these resonances as the \emph{interior} and \emph{exterior} resonances, respectively.

Adapting this idea for the general three-body problem, we analyze the Laplacian approximation of the Hamiltonian ${\mathcal H}$ (\ref{eq:ham0}-\ref{eq:ham1}), in order to compare the magnitudes of the terms related to the resonant angles $\sigma_1$ and $\sigma_2$. 
The first-order analytical expansion of the Hamiltonian  near the 3/1 MMR can be written as

\begin{eqnarray}
 H & = & A\left(I_1+I_2\right)+ 2B\left(I_1+I_2\right)^2 +   C I_1 + D I_2 + E \sqrt{I_1 I_2}\cos{\Delta\varpi}\nonumber\\
   & + & F_1 I_1 \cos{2\sigma_1} + F_2 I_2\cos{2\sigma_2} + F_3 \sqrt{I_1 I_2}\cos{\left(\sigma_1+ \sigma_2\right)},
 \label{hamiltoniano_laplace31}
\end{eqnarray}
where the terms with the coefficients $A$, $B$, $C$, $D$ and $E$ are associated to the secular mode of motion, while the terms with the coefficients $F_1$, $F_2$ and $F_3$ are associated to the resonant mode. The explicit expressions for the coefficients are given in Appendix.

Analyzing the resonant part of the Hamiltonian (\ref{hamiltoniano_laplace31}), we can observe that the term with $F_3$ is symmetric with respect to the angles $\sigma_1$ and $\sigma_2$ and thus does not constrain the choice between them. The comparison must be done between two other resonant terms, with coefficients $F_1$ and $F_2$, and the term with higher amplitude will indicate the resonant angle.

We apply the described above criterion to the systems in the nominal 3/1 MMR,  with different values of the mass ratio $m_2/m_1$ and the initial eccentricity $e_2$ and show the  results in Figure \ref{fig:f1f2f3_laplace}, where the condition $|F_2|-|F_1|>0$ is plotted by black symbols, while the condition $|F_2|-|F_1|<0$ is plotted by red symbols. The values of the coefficients $F_1$ and $F_2$ and are calculated using the orbital elements of the ($\pi$,$\pi$)--ACRs, for different values of the mass ratio. 
Figure \ref{fig:f1f2f3_laplace} shows that the choice of the resonant angle is determined essentially by the value of the ratio of the planetary masses: for $m_2/m_1$ larger than approximately 1.5, the 3/1 MMR is an interior resonance characterized by the libration of the critical angle $\sigma_1$ and oscillation/circulation of the secular angle $\Delta\varpi$ (and $\sigma_2$). For  $m_2/m_1$ lesser than 1.5, the 3/1 MMR is exterior, with the librating $\sigma_2$ and circulating $\Delta\varpi$ (and $\sigma_1$).  The influence of the planetary eccentricities on the choice of the critical angle seems to be negligible, at least, at first-order approximation.

It should be stressed that both angles, secular and resonant, are oscillating in the close vicinity of the stable ACR solutions. However, the oscillation of the secular angle in this case should be differentiated from the regime of motion defined as \textit{a libration} of the resonant angle. Indeed, the topology of the secular phase space defines a family of concentric curves around one center displaced from the origin; the curves which are close to the center, do not enclose the origin - they correspond to oscillations, while the outer curves enclose the origin and correspond to circulations. The separation between them is not a dynamical separatrix, characteristic of the resonant geometry,  but just one curve passing through the origin. The whole set of curves forms a homeomorphic family of solutions and the distinction between oscillations and circulations in this case is merely kinematical. To stress this behavior we refer to it as oscillatory/circulatory, when describing the secular regime of
motion (Michtchenko et al. 2011).

Finally, the location of the HD60532 planets indicated by a cross symbol on the ($m_2/m_1$,$e_2$)--plane in Figure \ref{fig:f1f2f3_laplace} shows that the system evolves in the internal 3/1 MMR and the critical angle $\sigma_1$ is a resonant angle.

\section{Topology of the semi-analytical Hamiltonian}\label{sec-3}

The analysis of the topology of the two-degrees-of-freedom Hamiltonian ${\mathcal H}$ (\ref{eq:ham0}-\ref{eq:ham1}) consists of several steps, such as the choice of the representative planes,  mapping the energy levels of ${\mathcal H}$ (\ref{eq:ham0}-\ref{eq:ham1}), the determination of the stationary solutions of the Hamiltonian function and the analysis of the stability of these solutions.


\subsection{Energy levels maps on the representative planes}\label{sec-3-1}
\label{repplane}

\begin{figure}
  \centering
  \includegraphics[width=1.0\textwidth]{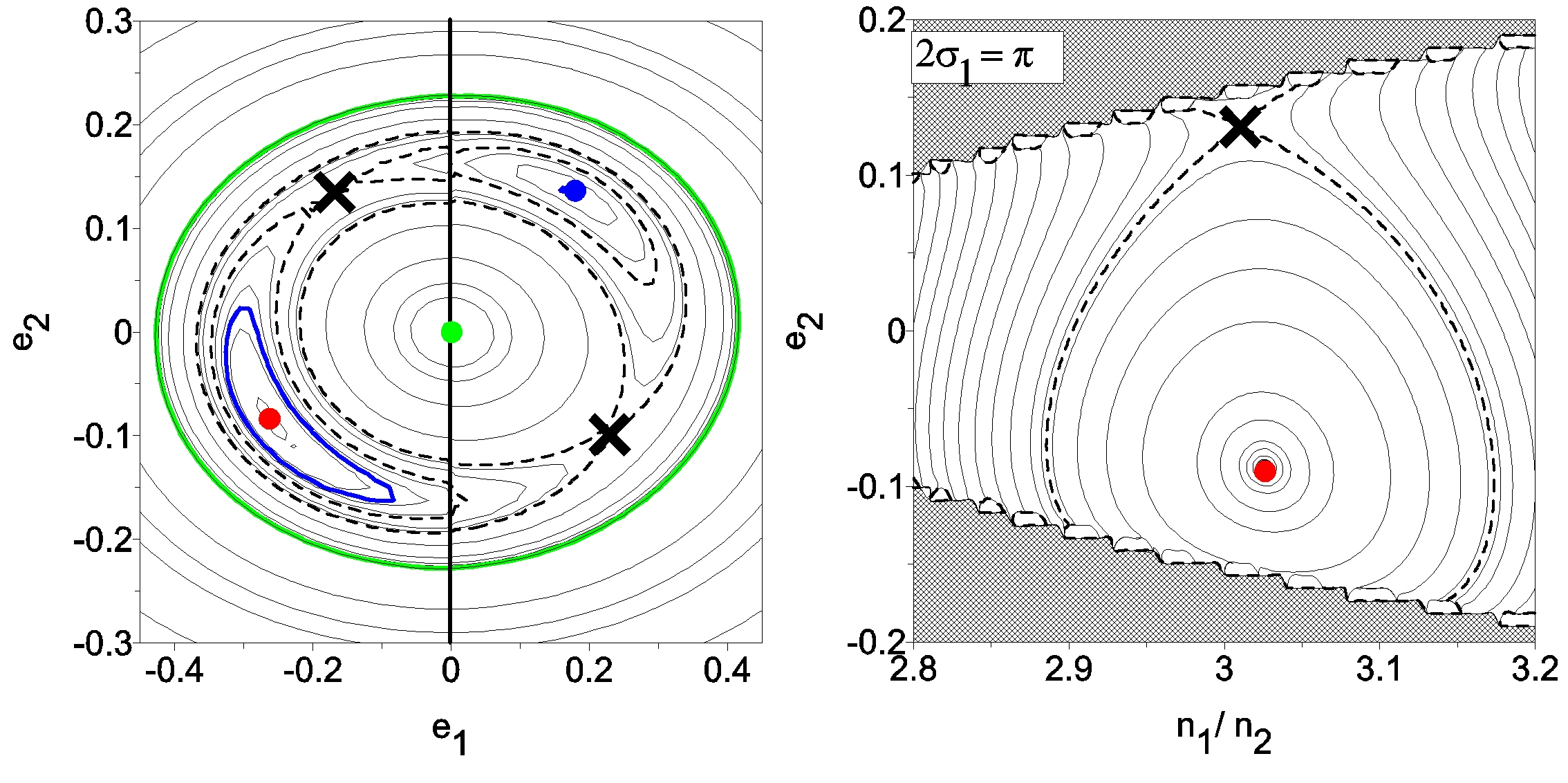}\\
  \caption{Energy levels of the 3/1 resonant Hamiltonian (\ref{eq:ham0}-\ref{eq:ham1}) on the ($e_1$,$e_2$) representative plane (left)  and on the ($n_1/n_2$, $e_2$)--plane obtained at $2\sigma_1=\pi$ (right). The positive (negative) values of $e_1$ indicate that $2\sigma_1=0 (\pi)$, while the positive (negative) values of $e_2$ indicate that $\Delta\varpi=0 (\pi)$. Both maps are constructed with the parameters ${\mathcal K}$ and ${\mathcal AM}$ corresponding to the HD60532 system. The symbols show locations of the stationary solutions of the Hamiltonian (see text for details).}
  \label{fig:phase_space123}
\end{figure}

Although the phase space of the two-degrees-of-freedom resonant system is four dimensional, the problem can be reduced to the systematic study of initial conditions on a plane. For instance, if we fix the initial values of the angular variables $2\sigma_1$  and $\Delta\varpi$, the phase space can be visualized on the ($e_1$, $e_2$)--plane of initial conditions. In this case, the semimajor axes of the planets (needed to calculate energy levels) may be obtained using Eqs. (\ref{spacing_parameter}-\ref{amtotal}), for a fixed set of the free parameters ${\mathcal K}$ and ${\mathcal AM}$. The representative plane ($e_1$, $e_2$), corresponding to the physical and orbital parameters of the HD60532 planets (Table \ref{Table:hd60532bc_data}), is shown in Figure \ref{fig:phase_space123}\,{\it left}, where the positive (negative) values on the $e_1$-axis correspond to $2\sigma_1$ fixed at $0$ ($180^\circ$), while  the positive (negative) values on the $e_2$-axis correspond to $\Delta\varpi$ fixed at $0$ ($180^\circ$).

Energy levels of the Hamiltonian ${\mathcal H}$ plotted on the ($e_1$, $e_2$)--plane show a complex structure of the 3/1 MMR, with several fixed points marked by symbols: the elliptic points by dots, while the saddle-type points by crosses. Each fixed point corresponds to a stationary solution of the averaged resonant problem, frequently referred to as  \emph{Apsidal Corotation Resonance} or ACR. The global maximum of ${\mathcal H}$ shown by a red dot in Figure \ref{fig:phase_space123}\,{\it left}, is located in the quadrant corresponding to $2\sigma_1=\Delta\varpi=180^\circ$, and is usually known as a $(\pi,\pi)$-ACR. A second maximum (blue dot) is found in the quadrant with $2\sigma_1=\Delta\varpi=0$ and characterizes the $(0,0)$-ACR solution; the corresponding energy level around the global maximum is shown by blue color.  Finally, there are two additional saddle-type fixed points shown by cross symbols in Figure \ref{fig:phase_space123}\,{\it left}.

It is expected that each elliptic stationary solution should be surrounded by a robust domain of stable resonant motion. This is shown in Figure \ref{fig:phase_space123}{\it right}; the figure shows  the energy levels  in the neighborhood of the global maximum (red dot) on the representative plane ($n_1/n_2$, $e_2$), where $n_1$ and $n_2$ are the mean motions of the inner and outer planets, respectively. Since the location of the global maximum is already known from the previous plane, we can fix the angle $2\sigma_1$ at $180^\circ$ and set $\Delta\varpi$ at $0$ (positive values on the $e_2$-axis) or $180^\circ$ (negative values on the $e_2$-axis). The initial values of $a_1$, $a_2$ and $e_1$ are then obtained from Eqs. (\ref{spacing_parameter}-\ref{amtotal}), using the same set of the free parameters ${\mathcal K}$ and ${\mathcal AM}$. The separatrix-like energy level surrounding the global maximum (red dot) is shown by dashed curve in Figure \ref{fig:phase_space123} {\it right}. Finally, the regions of the
($n_1/n_2$, $e_2$)--plane, where there are no solutions of Eqs. (\ref{spacing_parameter}-\ref{amtotal}) for the given constants, are shown by gray color. This representative plane will be widely used in the next sections to portray the main features of the 3/1 resonant dynamics.

We have assessed the stability of the second elliptic point, associated to the $(0,0)$-ACR and marked by a blue dot in Figure \ref{fig:phase_space123}\,{\it left}, applying the procedure described in Section \ref{sec-3-3}. The results have shown that this point is unstable at small and moderate eccentricities (see Figure \ref{fig:families}); therefore, we will not discuss the dynamics around it in this paper.


\subsection{Stationary orbits of the averaged 3/1 resonance problem (ACR)}\label{sec-3-2}

As shown in the previous section, ACRs appear as fixed points, of elliptic or saddle types, in the energy levels maps constructed for given values of the parameters ${\mathcal AM}$ and ${\mathcal K}$. To obtain these special solutions over a large range of the parameters, we employ the geometrical method presented in Michtchenko et al. (2006). As shown in that paper, the position and stability of the ACR solutions are only marginally dependent on the spacing parameter ${\mathcal K}$, thus, in this section, we fix ${\mathcal K}$ and vary the other two parameters, $m_2/m_1$ and ${\mathcal AM}$. Choosing the values of the semimajor axes $a_1$ and $a_2$ from Table \ref{Table:hd60532bc_data}, the value of the spacing parameter ${\mathcal K}$ is then obtained using Eq. (\ref{spacing_parameter}), for a given mass ratio.

Usually, ACRs are classified in two types: symmetric and asymmetric solutions (e.g. Beaug\'e et al. 2003, Lee 2004, Voyatzis \& Hadjidemetriou 2005). The symmetric solutions are characterized by stationary values of both resonant and secular angles at zero or $180^\circ$, while for asymmetric solutions the values of the angles are different from zero or $180^\circ$.


\subsubsection{Symmetric ACRs}\label{sec-3-3}

\begin{figure}
  \centering
  \includegraphics[width=0.8\textwidth]{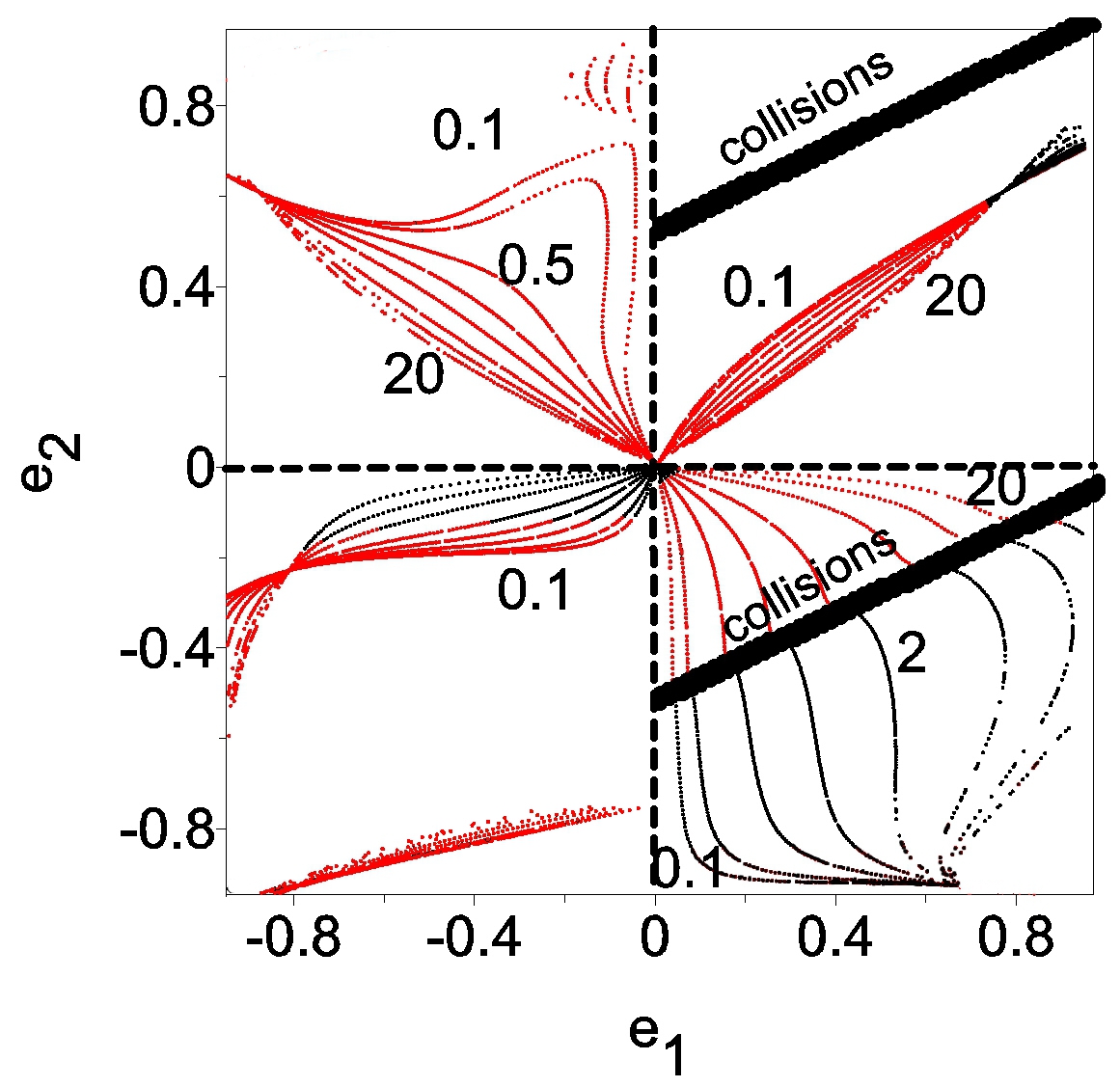}\\
  \caption{Families of symmetric ACRs of the 3/1 MMR, parameterized by the mass ratio $m_2/m_1$ $\in$ [0.1,20]. The positive (negative) values of $e_1$ correspond to $2\sigma_1$ equal to 0 ($\pi$), while the positive (negative) values of $e_2$ correspond to $\Delta\varpi$ equal to 0 ($\pi$). \emph{Black curves} correspond to stable solutions. \emph{Red curves} show unstable symmetric ACRs and \emph{broad lines} show locations of the initial conditions for which the planetary orbits become crossing, allowing collisions to occur.}\label{fig:families}
\end{figure}

\begin{figure}
  \centering
  \includegraphics[width=1.0\textwidth]{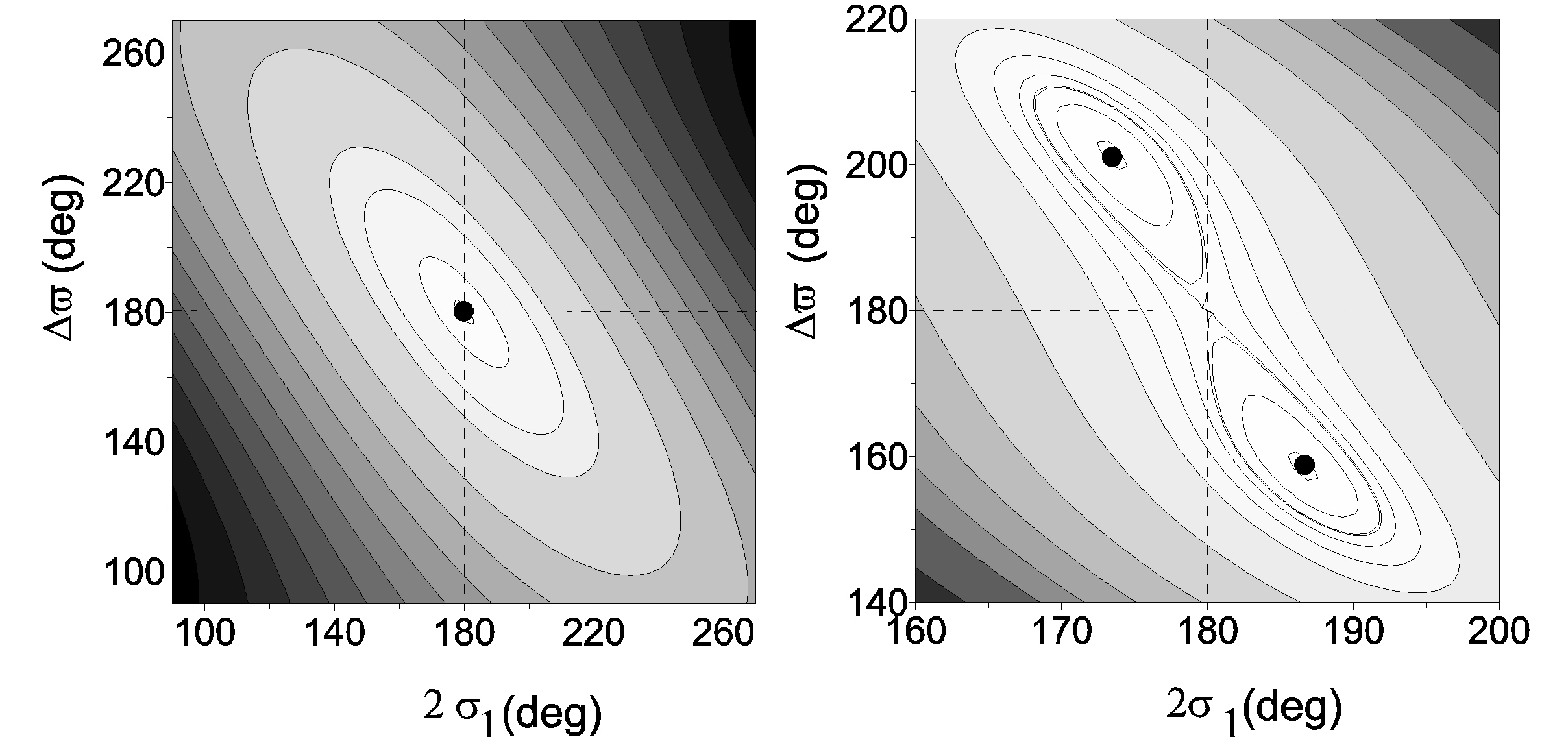}\\
  \caption{Level curves of the Hamiltonian (\ref{eq:ham0}-\ref{eq:ham1}) as functions of the resonant angle $2\sigma_1$ and the secular angle $\Delta\varpi$, around two symmetric ACRs with $m_2/m_1=2.36$. Left: Stable symmetric solution for $e_1 = 0.093$ and $e_2 = 0.035$. Right: Unstable symmetric solution for $e_1 = 0.45$ and $e_2 = 0.13$.}
  \label{fig:h_stab_1}
\end{figure}

Figure \ref{fig:families} shows the families of the symmetric 3/1 ACRs parameterized by the values of the mass ratio $m_2/m_1$, from 0.1 to 20, on the representative ($e_1$,$e_2$)--plane. Once again, the positive (negative) values on the $e_1$-axis correspond to $2\sigma_1$ fixed at $0$ ($180^\circ$), while  the positive (negative) values on the $e_2$-axis correspond to $\Delta\varpi$ fixed at $0$ ($180^\circ$). The thick lines are the locus of configurations leading to possible collisions between the planets and given by the condition
\begin{equation}
 a_1\left[1+e_1 \cos2\sigma_1\right]=a_2\left[1+e_2 \cos\Delta\varpi\right].\label{colision}
\end{equation}

The symmetric ACRs exist for all possible combinations of the angular variables, such as ($0$, $0$), ($\pi$, $0$), ($0$, $\pi$) and ($\pi$, $\pi$), where the numbers inside the brackets indicate the stationary values of $2\sigma_1$ and $\Delta\varpi$. However, not all of these solutions are stable. The stability of ACRs can be determined analyzing the structure of the Hamiltonian (\ref{eq:ham0}-\ref{eq:ham1}) in the vicinity of ACRs on the plane of the angular variables ($2\sigma_1$, $\Delta\varpi$).

An example is shown in Figure \ref{fig:h_stab_1}, where the energy levels are plotted by solid curves, while the background gray scale varies from dark tones to light ones, for increasing values of ${\mathcal H}$. The left graph shows a stable symmetric ($\pi$, $\pi$)--ACR (black dot) as an elliptic point of the maximal energy, while the right graph shows a saddle-type structure of the phase space, with an unstable symmetric ($\pi$, $\pi$)--ACR and two asymmetric elliptic points of the maximal energy (black dots).

Performing this analysis for all solutions shown in Figure \ref{fig:families}, we separate them into the stable (black dots) and unstable (red dots) ones. We find that the families of the stable symmetric ACRs corresponding to the global maximum of the Hamiltonian (\ref{eq:ham0}-\ref{eq:ham1}) are located in the $(\pi,\pi)$--quadrant in Figure \ref{fig:families}. The stable solutions can be found also in the $(0,\pi)$--quadrant: they have very high eccentricities and located beyond the collision curve. It is worth noting that the very-high-eccentricity ACRs in the $(0,0)$--quadrant become stable, but their counterparts from the $(\pi,0)$ and $(\pi,\pi)$--quadrants are always unstable. Although the dynamics of such eccentric planetary systems merits close scrutiny, in this paper, we focus on the study of the dynamics only around low-to-moderate-eccentricity $(\pi,\pi)$--ACR solutions.


\subsubsection{Asymmetric ACRs}

\begin{figure}
  \centering
  \includegraphics[width=0.8\textwidth]{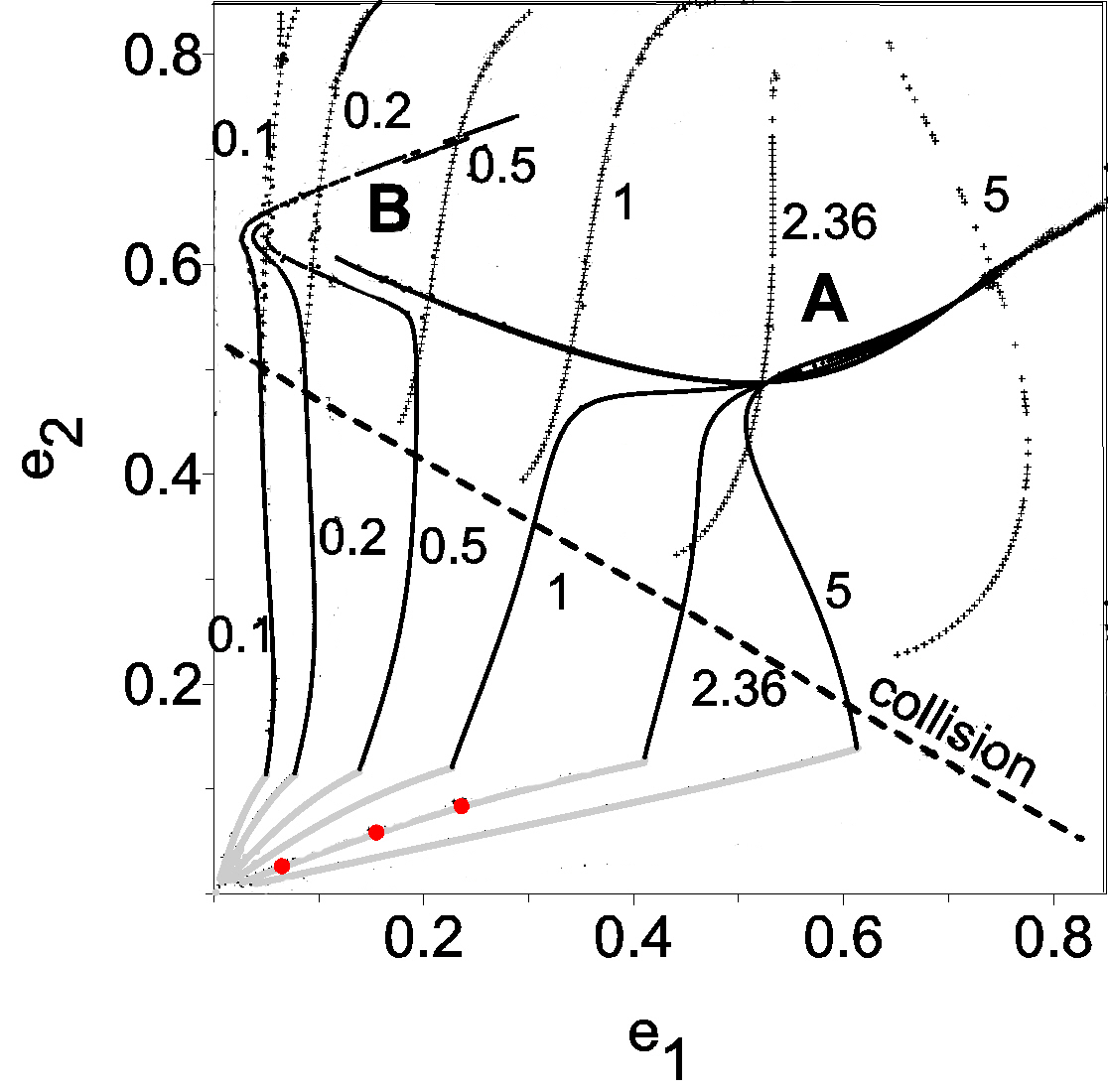}\\
  \caption{Families of the stable ACRs for the 3/1 MMR, parameterized by the mass ratio $m_2/m_1$. \emph{Gray solid lines} correspond to symmetric ($\pi$,$\pi$)--type ACRs, \emph{black solid lines} to stable asymmetric solutions, and \emph{black dots} to symmetric ($0$,$\pi$)--type ACRs. Dashed line shows the collision curve for $(0,\pi)-ACR$. Three red dots indicate positions of ACRs, whose domains will be mapped in Sect. \ref{dyn_out_acr}.}\label{fig:assymetric_1}
\end{figure}

\begin{figure}
  \centering
  \includegraphics[width=0.8\textwidth]{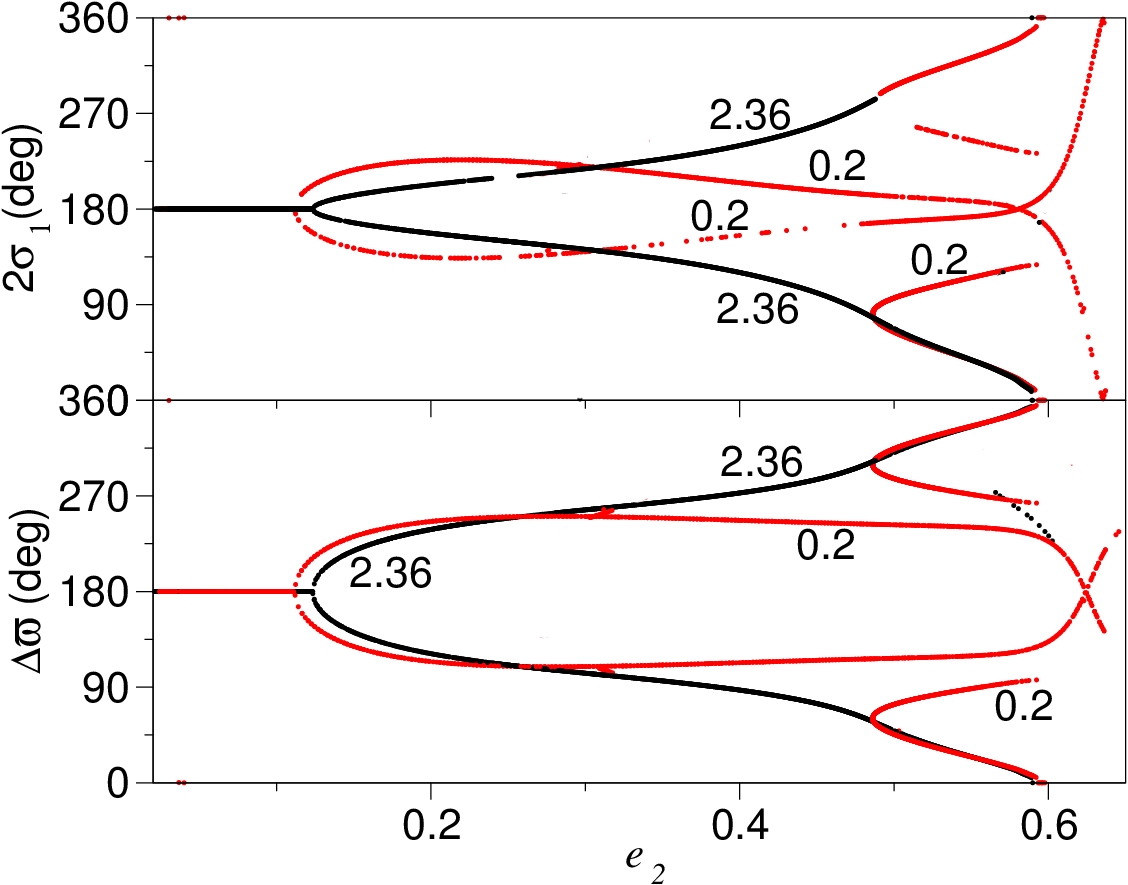}\\
  \caption{Stationary values of the resonant angle $2\sigma_1$ (top) and the secular angle $\Delta\varpi$ (bottom) for the two families of the ACRs, one parameterized by $m_2/m_1=2.36$ (black) and other by $m_2/m_1=0.2$ (red). Angles are shown as functions of the outer planet's eccentricity.}\label{fig:assymetric_2}
\end{figure}

The stable asymmetric ACRs of the 3/1 MMR appear at bifurcations of symmetric solutions along the ($\pi$,$\pi$)--families, when these pass from stable to unstable. This case is illustrated in Figure \ref{fig:h_stab_1}, right panel, where the hyperbolic ($\pi$,$\pi$)--point gives the origin to an asymmetric solution represented by two points. The calculation of the asymmetric ACRs adopts the geometrical approach presented in Michtchenko et al. (2006).

Figure \ref{fig:assymetric_1} summarizes the information on the stable symmetric and asymmetric ACRs of the 3/1 MMR showing their families parameterized by the mass ratio $m_2/m_1$ on the ($e_1$, $e_2$)--plane. At small eccentricities, all solutions are symmetric and of ($\pi$,$\pi$)--type; they are continuously increasing with the increasing planetary eccentricities (gray curves). This smooth evolution along symmetric families is interrupted by a sudden increase of the outer planet's eccentricity. The corresponding symmetric solutions become unstable and the stationary system evolves now along an asymmetric segment of the corresponding family (black curves).

For higher eccentricities, the families of asymmetric ACRs can bifurcate into two branches (identified as A and B in Figure  \ref{fig:assymetric_1}). To the left, we note a funneling of all solutions toward a narrow diagonal region (branch B).  To the right, we see a convergence point located approximately at ($e_1$,$e_2$) $= (0.5, 0.5)$, after which the solutions for all mass ratios practically coincide (branch A). Shortly following this convergence, the asymmetric solutions return to symmetric ACRs, this time of the (0,0)--type. A similar two-branch structure was also noted for the 2/1 MMR and was explained by effects of the true secular resonance \emph{inside} the mean-motion resonance (see Michtchenko et al. 2008b). The asymmetric solutions exhibit a very complex dynamics, which is out of the scope of this paper. Finally, the stable ($0$,$\pi$)--ACRs located beyond the collision curve are shown by dot symbols in Figure \ref{fig:assymetric_1}.

By definition, asymmetric ACRs are characterized by stationary values of both resonant and secular angles different from zero or $180^\circ$. The appearance and the evolution of the asymmetric solutions with the increasing outer planet's eccentricity is illustrated in Figure \ref{fig:assymetric_2}, for two different values of the mass ratio, 0.2 (red curves) and 2.36 of the HD60532 system (black curves). The angles $2\sigma_1$ and $\Delta\varpi$ are trapped at $180^\circ$  at small eccentricities that indicates symmetric character of the ACRs. The bifurcations to the asymmetric solutions occur around $e_2 \sim 0.1$, for both values of the mass ratio. The systems evolve in the asymmetric configurations up to $e_2=0.6$, when they return to symmetric configurations of the (0,0)--type.


\subsection{The Law of Structure}

\begin{figure}
  \centering
  \includegraphics[width=0.8\textwidth]{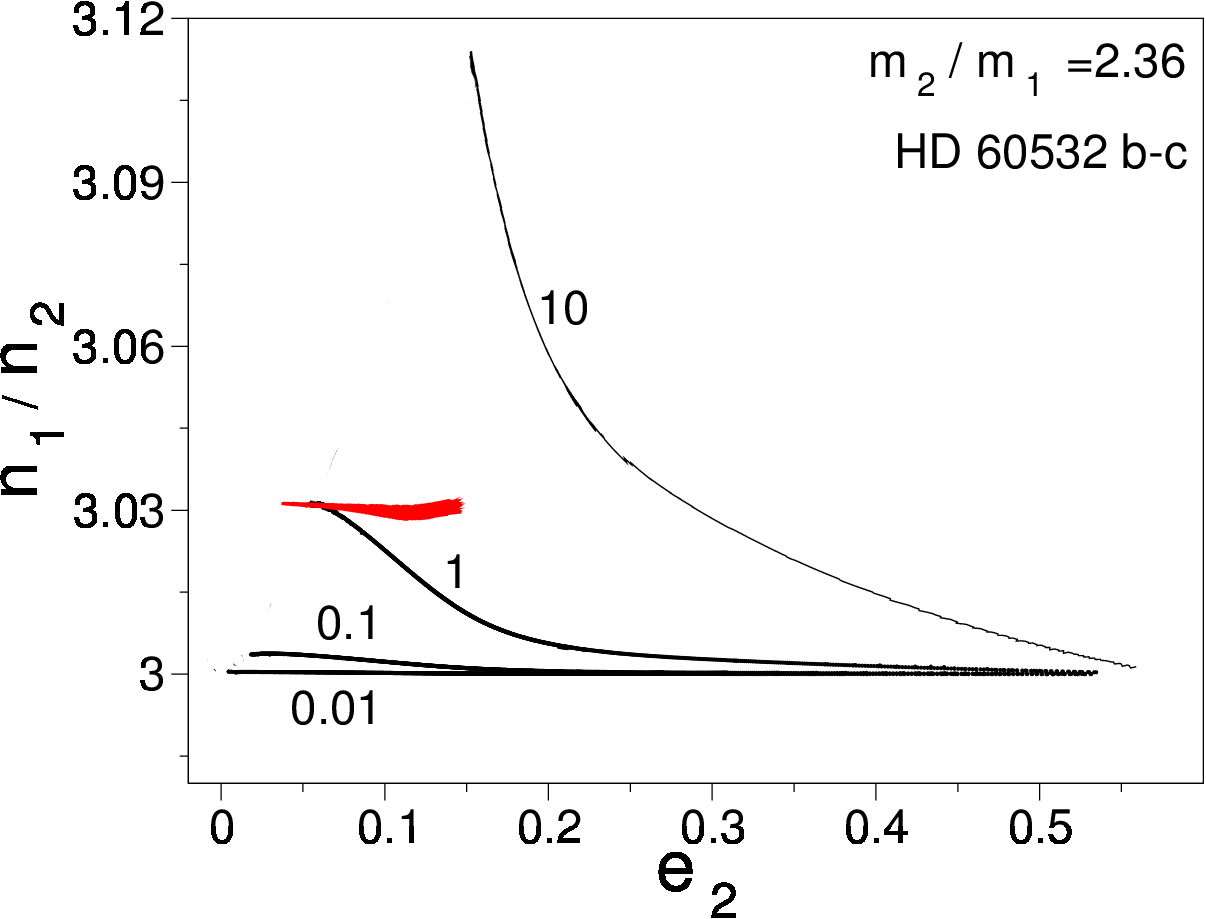}\\
  \caption{Variation of the ratio of planetary mean motions $n_1/n_2$ for symmetric $(\pi,\pi)$--type ACRs, as a function of the outer planet's eccentricity (Law of Structure). All curves are calculated for the fixed value of $m_2/m_1=2.36$ corresponding to the HD60532 planets, but  the values of the individual planetary masses are simultaneously multiplied by the factor whose values are shown beside each curve. \emph{Red curve} is a projection of the HD60532 resonant oscillation on the ($e_2$,$n_1/n_2$)--plane.
}\label{fig:law_structures_simplex_2}
\end{figure}

It is important to emphasize that the ratios of the  mean  motions of the resonant stationary configurations of the planets generally deviate from the nominal value 3/1 and are functions of the free parameters, ${\mathcal K}$ and ${\mathcal AM}$ (i.e. of the eccentricities of the ACRs). This feature is known as the 'Law of Structure' and was first observed in the studies of the asteroidal resonant dynamics (Ferraz-Mello 1988).

Figure \ref{fig:law_structures_simplex_2} shows the evolution of the mean-motion ratio of two planets in the stationary configuration as a function of the  eccentricity of the outer planet. In all cases, the masses of the planets were fixed at values of the masses  of the HD60532 system, with $m_2/m_1 = 2.36$, but we adopted several values of the individual planetary mass factor, which are indicated beside each curve in Figure \ref{fig:law_structures_simplex_2}. The variation of $n_1/n_2$ with the eccentricity  shows that, for stationary solutions, the ratio of orbital periods deviates from $3/1$, in such a way that its value is always above the nominal resonance value. Therefore, resonance pairs evolving in nearly-stationary configuration (that is, with small oscillations) will be always concentrated at one side of the nominal value 3/1, for instance, at the left side in Figure \ref{fig:planetary_systems31}. Moreover, the deviation will increase for the decreasing values of eccentricities.  The
projection
of the oscillation of the HD60532 planetary pair around the corresponding stationary solution (ACR) is shown by red dots in Figure \ref{fig:law_structures_simplex_2}. The resonant oscillations of the semimajor axes of the planets shown in Figure \ref{fig:aiei_oscillation} appear now as a small oscillation of the mean-motion ratio around  $n_1/n_2 \sim  3.03$, while the outer planet's eccentricity varies in the range between 0.05 and 0.15.

Figure \ref{fig:law_structures_simplex_2} also shows that the mean-motion ratio is sensible to values of the individual planetary masses. Indeed, the deviation from the nominal value $3/1$ is an increasing function of the increasing individual planetary masses, even if the mass ratio is kept fixed. This feature contrasts with the behaviour of the stationary values of eccentricities and angular variables, which are only marginally dependent on the individual planetary masses (Beaug\'e \& Michtchenko 2003).


\section{Dynamics around the ACRs}\label{dyn_out_acr}

To analyze the dynamics of the 3/1 resonance in the region of the phase space beyond each symmetric ACR, we apply Spectral Analysis Method (SAM) developed in Michtchenko et al. (2002) and described in details in Ferraz-Mello et al. (2005). The method allows us to study the structure of the phase space around ACRs in form of dynamical maps on the $(n_1/n_2, e_2)$ representative planes of initial conditions.

Figures \ref{fig:mapa_dinamico_dynamic_power_spectra_e1_0.06}, \ref{mapa_dinamico_dynamic_power_spectra_e1_0.15} and \ref{mapa_dinamico_dynamic_power_spectra_e1_0278} show the dynamical maps constructed around three different values of ACR chosen along the family parameterized by $m_2/m_1=2.36$ (see Figure \ref{fig:assymetric_1}). It is worth emphasizing that the topology of the maps is uniquely defined by two free parameters of the problem, ${\mathcal K}$ and ${\mathcal AM}$. ${\mathcal K}$  is same for all maps; its value is fixed at 0.6345  (in units of the solar mass, astronomical unit and year), and corresponds to the physical and orbital parameters of the HD60532 system in Table \ref{Table:hd60532bc_data}.

The value of the angular momentum ${\mathcal AM}$ varies from one map to other. In this paper, instead of giving this value, we match each solution with the value of the eccentricity of the inner planet. In this way, it is easy to identify the location of the studied ACRs shown by three red dots on the corresponding family in Figure \ref{fig:assymetric_1}. The ACRs under study characterize three different ranges of the planetary eccentricities: low ($e_1=0.06$), moderate ($e_1=0.15$) and high ($e_1=0.278$), the last value corresponding to the current value  of the inner planet of the HD60532 system.


\subsection{General characteristics of dynamical maps}

Although many aspects of the dynamical maps change as functions of ${\mathcal AM}$, there are also some general features, which are common for different sets of the constants.

\begin{figure}
  \includegraphics[width=1.1\textwidth]{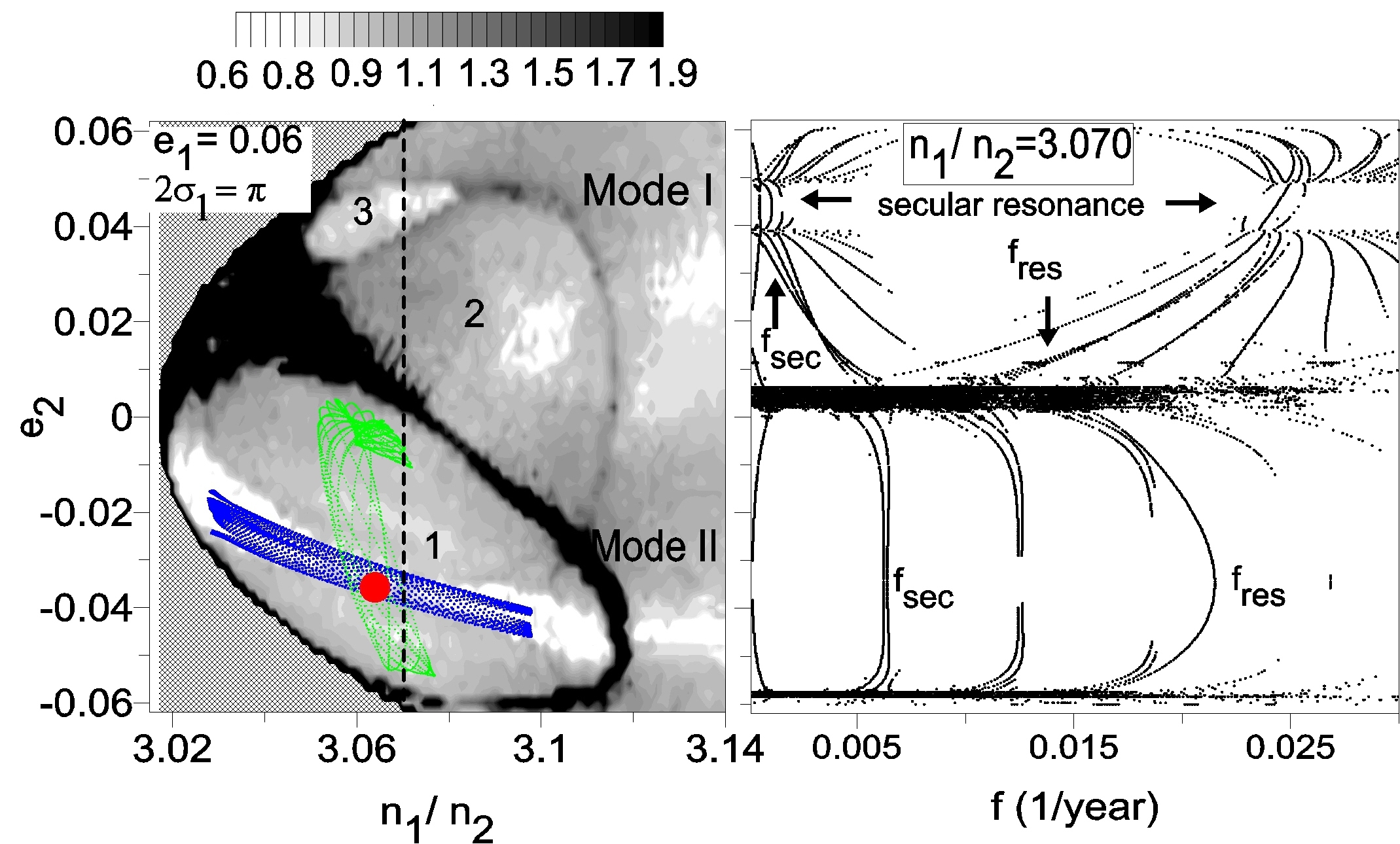}
  \caption{Left: Dynamical map of the domain around the stable symmetric ($\pi$,$\pi$)--ACR with $e_1=0.06$ and $m_2/m_1=2.36$. The initial value of $2\sigma_1$ is fixed at $180^\circ$ and the initial value of $\Delta\varpi$ is fixed at 0 (positive values on the $e_2$-axis) or $180^\circ$ (negative values on the $e_2$-axis)). \emph{Red circle} shows location of ACR. Number \textbf{1}, \textbf{2} and \textbf{3} indicate domains of distinct regimes of motion. \emph{Green} and \emph{blue dots} are projections of two planetary paths (see text for details). Right: Dynamical power spectrum obtained along the horizontal dashed line in dynamical map, showing the proper frequencies of the system as functions of $e_2$. The smooth evolution of the proper frequencies $f_{\textrm{res}}$ and $f_{\textrm{sec}}$ is characteristic of the regular motion, while the erratic scattering of the points is characteristic of the chaotic motion.}\label{fig:mapa_dinamico_dynamic_power_spectra_e1_0.06}
\end{figure}

Each map in Figures \ref{fig:mapa_dinamico_dynamic_power_spectra_e1_0.06}, \ref{mapa_dinamico_dynamic_power_spectra_e1_0.15} and \ref{mapa_dinamico_dynamic_power_spectra_e1_0278} contains a center, which represents the $(\pi,\pi)$--ACR solution; its position is shown by a red dot on the map. Around each center we find a region of quasi-periodic motion coded in gray scale which represents the spectral number $N$ (stability indicator) in logarithmic scale. The lighter tones for smaller values of $N$ correspond to regular motion, while the darker tones for higher values of $N$ correspond to increasingly chaotic motion. The spectral number was obtained through the Fourier analysis of the mean-motion ratio time series calculated over 100K years at each node of the grid of initial conditions; it is shown by color-scale bars in Figures \ref{fig:mapa_dinamico_dynamic_power_spectra_e1_0.06}, \ref{mapa_dinamico_dynamic_power_spectra_e1_0.15} and \ref{mapa_dinamico_dynamic_power_spectra_e1_0278}. We considered all 
frequencies peaks with at least 5\% of the maximum amplitude peak. The hatched areas are the region of forbidden motion  where there are no solutions of Eqs. (\ref{spacing_parameter}-\ref{amtotal}) for a given set of the constants ${\mathcal K}$ and ${\mathcal AM}$.

The quasi-periodic behavior inside the 3/1 MMR is an interaction of two modes of motion, associated with the two independent frequencies of the averaged resonant system. These modes are the resonant (with the frequency $f_{\textrm{res}}$) and the secular (with the frequency $f_{\textrm{sec}}$) modes of motion. The first is associated to the oscillation of the semimajor axes (see Figure \ref{fig:aiei_oscillation}\,\textit{top}) and resonant angle $2\sigma_1$. The second mode of motion is associated to the secular oscillation of the eccentricities (see Figure \ref{fig:aiei_oscillation}\,\textit{bottom}) and the difference in longitudes of pericenter, $\Delta\varpi$.

Any regular solution will be given by a linear combination of two periodic terms, each with a given amplitude and phase angle. Generally, both modes are well separated in the frequency space, with $f_{\textrm{res}}$ much higher than the secular frequency $f_{\textrm{sec}}$. However, at small eccentricities, both frequencies may have a same order, and it is possible to find initial conditions corresponding to low-order commensurabilities between them. These commensurabilities give origin to so-called {\it secondary resonances} inside the primary 3/1 resonance, which are also observed in the asteroidal dynamics (e.g. Moons and Morbidelli 1993).

The domains of chaotic motion are always present on dynamical maps of the 3/1 mean-motion resonance. The chaos is associated with the existence of separatrices between the 3/1 resonance region and the regions of quasi-resonant and purely secular motion, and between the secondary resonances inside the primary 3/1 resonance. The domains of strongly chaotic motion appear in  dark tones on the dynamical maps and are regions of large-scale instabilities followed by disruption of the system within the lifetime of the central star.

Finally, the dynamical maps in Figures \ref{fig:mapa_dinamico_dynamic_power_spectra_e1_0.06}, \ref{mapa_dinamico_dynamic_power_spectra_e1_0.15} and \ref{mapa_dinamico_dynamic_power_spectra_e1_0278} are always accompanied by dynamical power spectra, whose construction is described in detail in Michtchenko et al. (2002). Briefly, the spectra show the evolution of the main frequencies of the systems, $f_{\textrm{res}}$, $f_{\textrm{sec}}$ and their linear combinations,  as functions of either the mean-motion ratio $n_1/n_2$ or the eccentricity of the outer planet $e_2$.  The smooth evolution of the frequencies is characteristic of regular motion, while the erratic scattering of the frequency values is characteristic of the chaotic motion. In addition, the domains where one of the frequencies tends to zero accurately indicate the location of the separatrices between distinct regimes of motion.


\subsection{Low-eccentricity symmetric ACR dynamical maps}

The first map (Figure \ref{fig:mapa_dinamico_dynamic_power_spectra_e1_0.06}\,\textit{left}) was calculated for a small initial eccentricity of the inner planet ($e_1 = 0.06$) and illustrates the transition from the purely secular regime of motion towards the resonant one. Indeed, the capture of the migrating planetary pair into the 3/1 MMR is more likely when the planets evolve on the nearly circular orbits (see Figure \ref{fig:mig2_zeroecc}).

For initial conditions far from the center of the resonance (for instance, any point with $n_1/n_2>3.12$), the system is evolving in the purely secular regime. Its motion is defined by the composition of two normal secular modes, which are usually referred to as {\it Mode I} and {\it Mode II} of motion. Mode I corresponds to the aligned configuration of the planets ($\Delta\varpi=0$), with the minimal possible energy of the system for a given ${\mathcal AM}$. Mode II describes an anti-aligned configuration ($\Delta\varpi=180^\circ$) and corresponds to the maximal values of the energy. On the dynamical map, these modes appear as light-colored strips in the domain of the secular motion.

The modeling of the secular evolution of migrating planet pairs done in Michtchenko and Rodr\'iguez (2011), has shown that, during convergent migration, the planets converge and evolve along the Mode II, as long as $m_2/m_1>\sqrt{a_1/a_2}$ (see Figure 2 in that paper). This is the case of the HD60532 planet pair, thus the entrance of the planets into the 3/1 mean-motion resonance would occur through the Mode II of motion of the maximal energy, when the secular angle $\Delta\varpi$ oscillates around $180^\circ$. Numerical simulations of planetary migration confirm this result (e.g. Ferraz-Mello et al. 2003, Kley et al. 2005, Beaug\'e et al. 2006).

Figure \ref{fig:mapa_dinamico_dynamic_power_spectra_e1_0.06} (left panel) shows that the domain of the 3/1 MMR is separated into three distinct regions surrounded by layers of chaotic motion. These regions are associated with different regimes of motion of the system and we denote them by the numbers \textbf{1}, \textbf{2} and \textbf{3} in Figure \ref{fig:mapa_dinamico_dynamic_power_spectra_e1_0.06}. The dynamical power spectrum (right panel) constructed along the vertical line with $n_1/n_2=3.07$  shows that, during transitions between the different regimes, one of the independent frequencies, $f_{\textrm{res}}$ and $f_{\textrm{sec}}$, tends to zero. This feature is associated to the dynamics close to separatrix.

The dynamics in the regime \textbf{1} is a proper 3/1 MMR, with the resonant angle $\sigma_1$ librating and the secular angle $\Delta\varpi$ oscillating/circulating around $180^\circ$. The resonant motion is a composition of two independent modes, resonant and secular. To illustrate that, we show in Figure \ref{fig:mapa_dinamico_dynamic_power_spectra_e1_0.06}\,\textit{left} the projections on the ($n_1/n_2$,$e_2$)--plane of the two nearly periodic orbits. The green curve represents the low-amplitude oscillations in the resonant mode; therefore, in this case, we observe mainly the slow secular component in the orbital oscillations. The exchange of ${\mathcal AM}$ between the two planets affects only the eccentricities, while the semimajor axes are almost constant in this case. On contrary, the blue curve in Figure  \ref{fig:mapa_dinamico_dynamic_power_spectra_e1_0.06}\,\textit{left}  is a projection of the orbit which exhibits mainly the fast resonant oscillations, in such a way that $\Delta\varpi$ is almost
constant and $2\sigma_1$ librates with large amplitude around the ACR.

\begin{figure}
  \centering
  \includegraphics[width=1.0\textwidth]{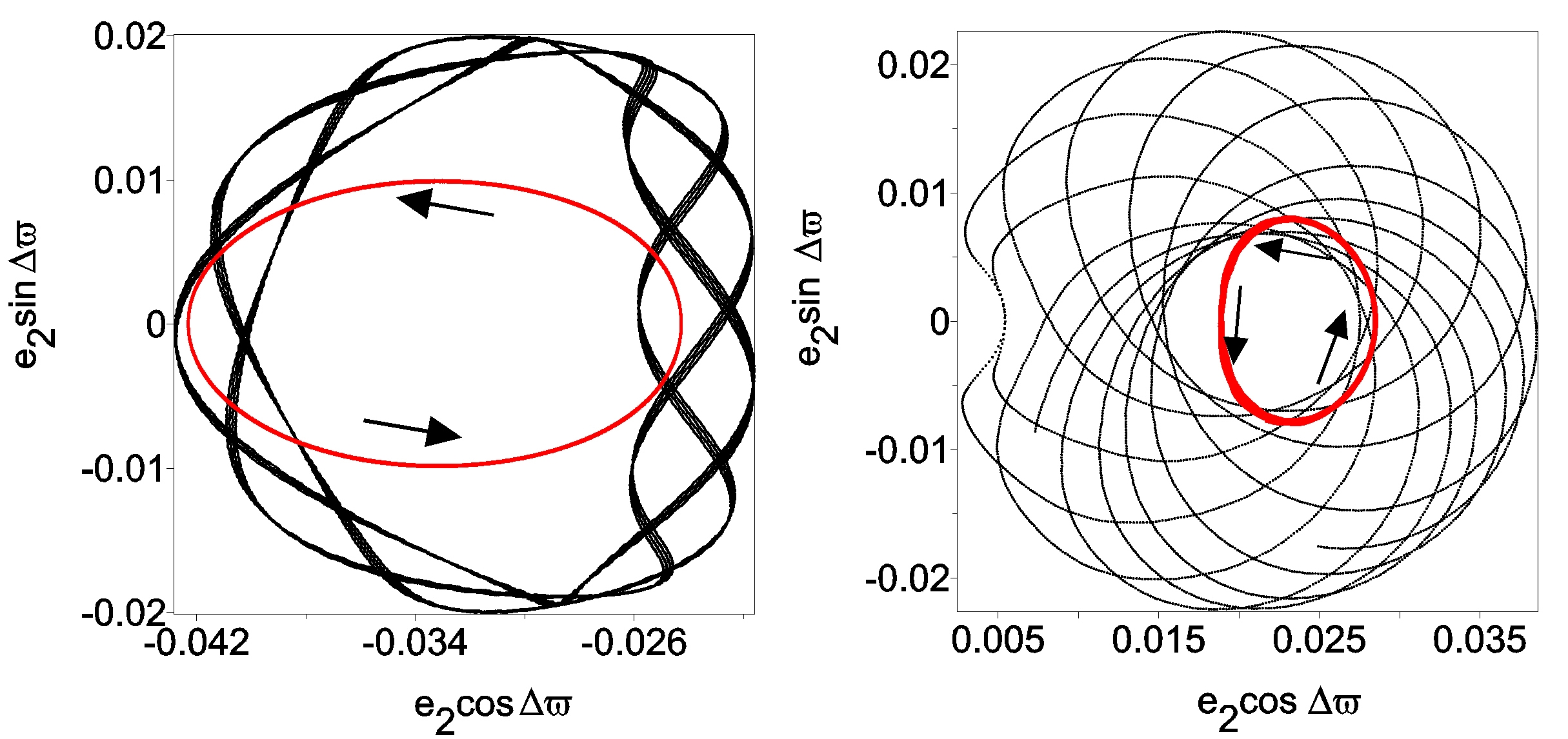}
  \caption{Left: \emph{Black line} shows one orbit from the domain \textbf{1} on the dynamical map in Figure \ref{fig:mapa_dinamico_dynamic_power_spectra_e1_0.06}\,left. \emph{Red line} is a secular component of this orbit. Right: Same as on the left panel, except for the domain \textbf{2}.}\label{fig:angles}
\end{figure}

\begin{figure}
  \centering
  \includegraphics[width=0.7\textwidth]{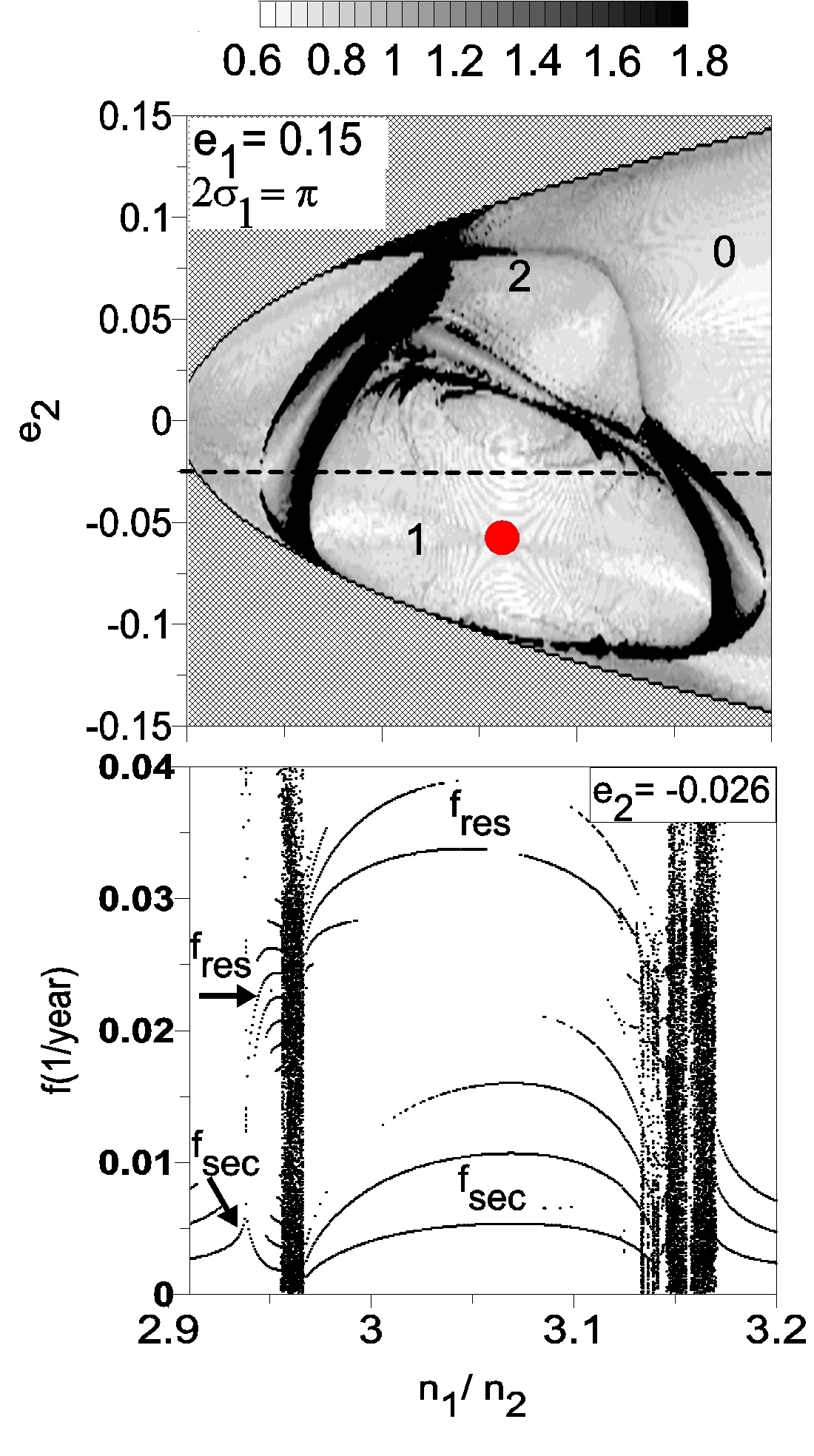}\\
  \caption{Top: Same as in Figure \ref{fig:mapa_dinamico_dynamic_power_spectra_e1_0.06}\,\textit{left}, except for $e_1=0.15$. Bottom: Same as in Figure \ref{fig:mapa_dinamico_dynamic_power_spectra_e1_0.06}\,\textit{right}, except calculated along the horizontal dashed line.}\label{mapa_dinamico_dynamic_power_spectra_e1_0.15}
\end{figure}

Figure \ref{fig:angles}\,\textit{left} shows separately the two components, resonant and secular, of a typical orbit in the regime \textbf{1} of motion. The red curve represents the secular mode and the sets show prograde direction of the oscillation around $180^\circ$. The black curve is a superposition of the resonant mode on the secular one. The typical values of the resonant and secular periods of oscillations can be assessed from the  dynamical power spectrum in Figure \ref{fig:mapa_dinamico_dynamic_power_spectra_e1_0.06}\,\textit{right}; they are of order of several tens and some hundreds years, respectively.

The domain of the regime \textbf{2} of motion is a quasi-resonant region characterized by the simultaneous circulation of both $2\sigma_1$ and $2\sigma_2$, in such way that the secular angle (which by definition is $\Delta\varpi=\sigma_1-\sigma_2$) oscillates around zero. A typical orbit from this region is shown in the ($e_2\cos\Delta\varpi$,$e_2\sin\Delta\varpi$)--plane by the black curve in Figure \ref{fig:angles}\,\textit{right}. Once again, we separate the secular component of the oscillation applying low-pass filtering; it  is shown by the red curve in Figure \ref{fig:angles}\,\textit{right}. Note that, although the secular oscillation occurs now around zero, it is still prograde, similar to one shown on the top graph of the same figure.

The distinct behaviour of the system in the regime \textbf{2} can be understood analyzing the evolution of the main frequencies on the dynamical power spectrum in Figure \ref{fig:mapa_dinamico_dynamic_power_spectra_e1_0.06}\,\textit{right}. The domain of this regime is located above the layer of strongly chaotic motion at $e_2\sim 0$, which separates the libration of the resonant angle $2\sigma_1$ from its circulation. The main feature of the dynamics in the regime \textbf{2} is a decay of the secular frequency $f_{\textrm{sec}}$ with increasing eccentricity of the outer planet, approaching zero at $e_2\sim 0.04$.  This is due to fact that, at this condition, the secular angle $\Delta\varpi$ changes the direction of its oscillation/circulation from prograde, in the quasi-resonant regime, to retrograde, in the purely secular regime of motion of the system. For some initial conditions, this transition raises a very specific regime of the stable motion, which we refer to as 'true secular resonance'. On
the dynamical map in \ref{fig:mapa_dinamico_dynamic_power_spectra_e1_0.06}\,\textit{left} the domain of this regime is marked by the number \textbf{3}.

The domain of the secular resonance is small; it is extended in the range $0.04 - 0.05$ along the eccentricity of the outer planet and is delimited by the narrows layers of chaotic motion associated with this resonance. The dynamics inside the secular resonance characterized by the libration of the secular angle $\Delta\varpi$ around zero, while the resonant angle $2\sigma_1$ is in retrograde circulation.

\subsection{Moderate-eccentricity symmetric ACR dynamical maps}

We present the dynamics  of the system at moderate eccentricities of the planets in Figure \ref{mapa_dinamico_dynamic_power_spectra_e1_0.15}.  On the top panel of the figure, we plot the dynamical map constructed for $e_1=0.15$, while, on the bottom graph, we present the dynamical power spectrum calculated along the $n_1/n_2$--axis, with $e_2$ fixed at $-0.026$. On the dynamical map, this $e_2$--value is represented by the dashed horizontal line.

\begin{figure}
  \centering
  \includegraphics[width=0.8\textwidth]{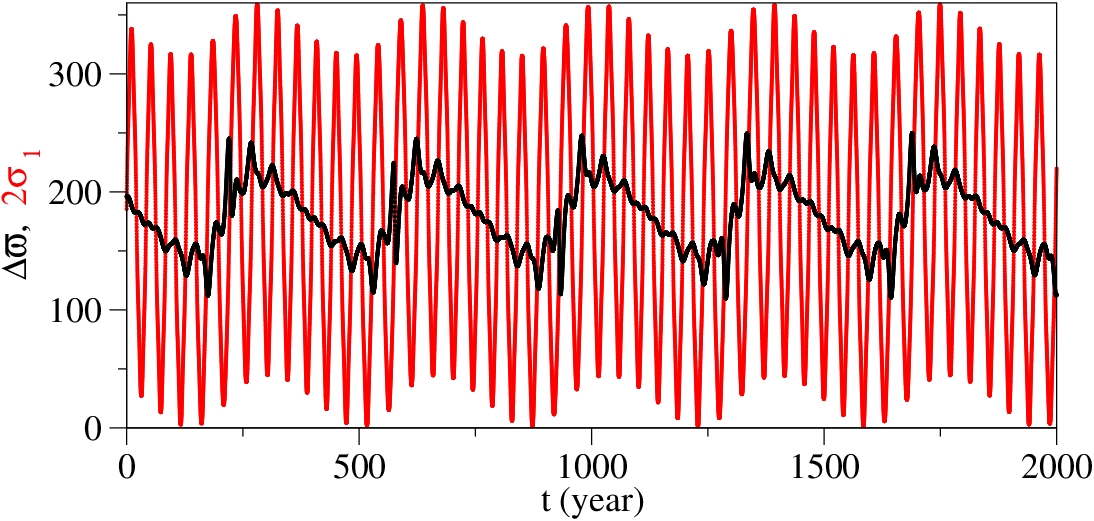}\\
  \caption{Stable evolution of the resonant angle $2\sigma_1$ (red) and the secular angle $\Delta\varpi$ (black), typical for the initial conditions inside the secular resonance. }\label{angles_behaviour_pi_2}
\end{figure}

Comparing to the low-eccentricity dynamics shown in Figure \ref{fig:mapa_dinamico_dynamic_power_spectra_e1_0.06}, we can note that the planetary dynamics shows no qualitative changes at increasing eccentricities. Both main regimes of motion, resonant and quasi-resonant, are present on dynamical map in  Figure \ref{mapa_dinamico_dynamic_power_spectra_e1_0.15}\,\textit{top}. As expected, the domain of the 3/1 MMR increases, while the layers of chaotic motion are expanded. The resonant regime of motion \textbf{1} is a combination of two oscillation modes, resonant and secular, around the corresponding ACR solution (red dot), located at $2\sigma_1=\Delta\varpi=180^\circ$. The evolution of the two independent  frequencies, $f_{\textrm{res}}$ and $f_{\textrm{sec}}$ can be observed on the dynamical power spectra in Figure \ref{mapa_dinamico_dynamic_power_spectra_e1_0.15}\,\textit{bottom}.

The quasi-resonant regime of motion \textbf{2} also preserves its main properties, when both critical angles $2\sigma_1$ and $2\sigma_2$, circulate retrogradely, while the secular angle oscillate/circulates in the prograde direction. The broaden chaotic layer separating the domain of the quasi-resonant motion from that of the purely secular motion does not allow us to detect the presence of the secular resonance in the regime \textbf{3}.

On the other hand, some new features appear in the dynamics at the moderate eccentricities in Figure \ref{mapa_dinamico_dynamic_power_spectra_e1_0.15}. One of these features are secondary resonances located at small eccentricities and forming the web of the periodic and chaotic orbits in the transition from the resonant \textbf{1} to the quasi-resonant \textbf{2} regimes of motion. The most strong of these is a $4/1$ commensurability between the frequencies $f_{\textrm{res}}$ and $f_{\textrm{sec}}$ represented by the family of periodic solutions inside the chaotic layer in this region (light tones).
\begin{figure}
  \centering
  \includegraphics[width=1.0\textwidth]{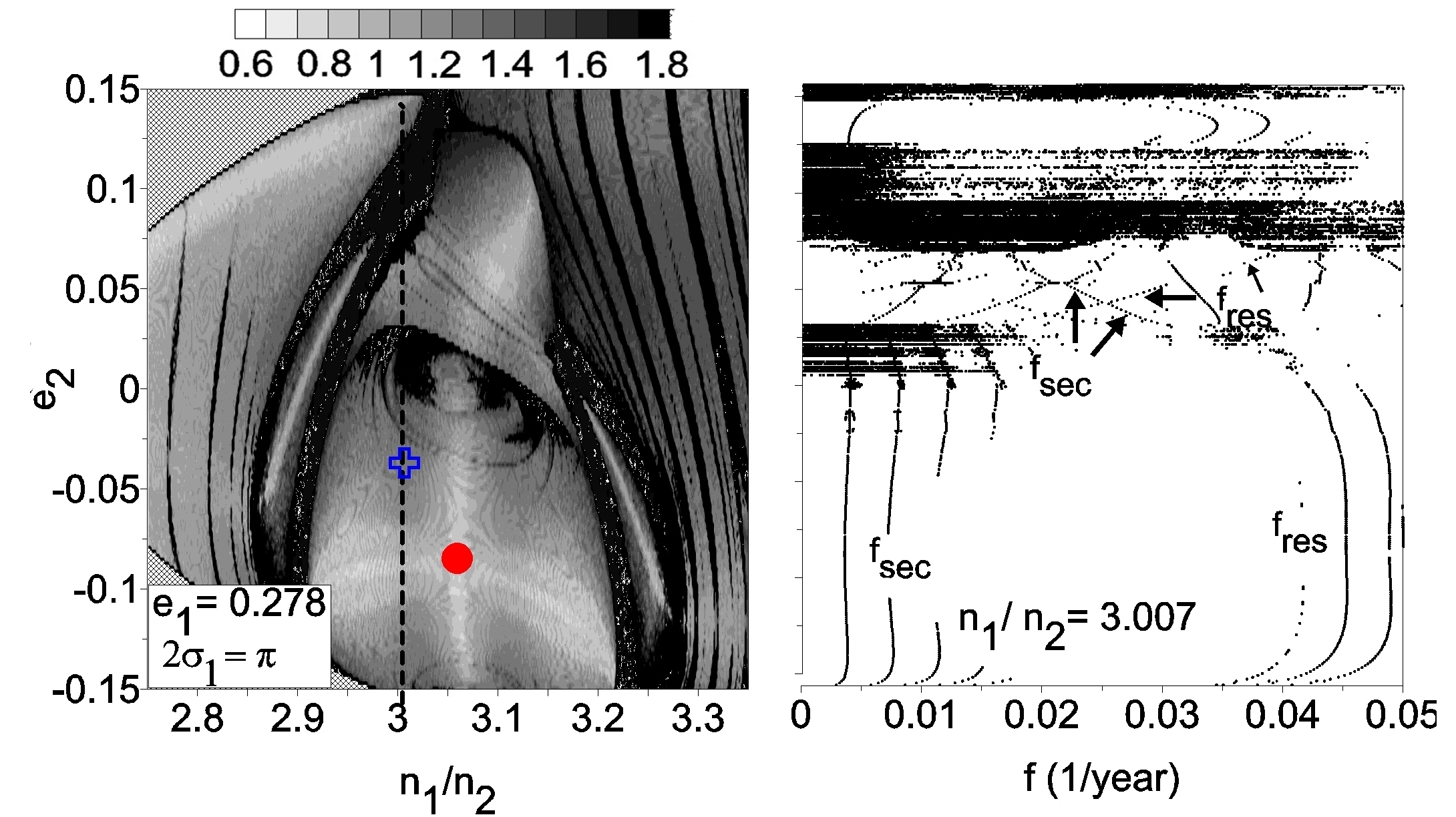}\\
  \caption{Same as in Figure \ref{fig:mapa_dinamico_dynamic_power_spectra_e1_0.06}, except for $e_1=0.278$. \emph{Blue symbol} shows the location of the HD60532 b-c system on the dynamical map.}
  \label{mapa_dinamico_dynamic_power_spectra_e1_0278}
\end{figure}

The other feature is a complex structure of the chaotic layers which separate the 3/1 MMR domain from the secular one. We can observe the existence of stable periodic solutions just inside the separatrices. Figure \ref{angles_behaviour_pi_2} shows a typical behaviour of the resonant and secular angles inside these narrow regions of stable motion. We note that, while the resonant angle $2\sigma_1$ librates around $180^\circ$ with very large amplitudes ($\sim 150^\circ$), the secular angle $\Delta\varpi$ oscillates around the same value, but with small amplitude. The similar property was also observed in the case of the $3/2$ and $2/1$ asteroidal resonance dynamics (Michtchenko \& Ferraz-Mello 1995) and was associated to the true secular resonance, which appears in the transition from the  resonant zone, where $\Delta\varpi$ oscillates/circulates progradely, to the secular zone, where the direction of the circulation of the secular angle is inverted.


\subsection{High-eccentricity symmetric ACR dynamical map}

The physical and orbital parameters of the HD60532 b-c  system, with $e_1=0.278$ (see Table \ref{Table:hd60532bc_data}), were used to present the high-eccentricity dynamics of the 3/1 MMR. Figure \ref{mapa_dinamico_dynamic_power_spectra_e1_0278} shows the corresponding dynamical map (left panel) and dynamical power spectrum (right panel). The spectrum was calculated along the dashed vertical line on the dynamical map, with the fixed value $n_1/n_2=3.007$, which corresponds to the HD60532 planets. The location of the system on the dynamical map is shown by a cross symbol and the red circle is a center of the resonant region where the HD60532 planets evolve.

The structure of the phase space observed in Figure \ref{mapa_dinamico_dynamic_power_spectra_e1_0278}\,\textit{left} is very similar to that observed at moderate eccentricities. The domain of the resonant motion surrounding the central ACR (red circle) is dominating on the dynamical map and is separated from the region of the quasi-resonant motion by the thick chaotic layer of the complex structure. The  HD60532 system is located inside the resonance zone and oscillates around the corresponding ACR, with the $2\sigma_1$--amplitude  of $\sim 120^\circ$ and the $\Delta\varpi$--amplitude  of $\sim 80^\circ$. Note that such large value of the amplitude of the critical angle is due to the composition of the resonant and secular modes in its oscillation, both with amplitudes of $\sim 60^\circ$.  The numerical tests have shown that the behaviour of the system is stable over timespan of 5G years (Laskar \& Correia 2009). However, it is worth emphasizing that the presence of the tiny chaotic layers associated to
secondary resonances in the close vicinity of the system could destabilize its motion in the case of possible corrections of the orbits due to future observations. Thus, to assess the long-term stability of the HD60532 system, the more detailed study of the domain where the system evolves, is needed.

The region of the quasi-resonant regime of motion shows a complex structure, with several strips of chaotic motion crossing the region. The analysis of the dynamical power spectrum in this region (right panel in Figure \ref{mapa_dinamico_dynamic_power_spectra_e1_0278}) confirms  that, in this regime of motion, the independent frequencies $f_{\textrm{sec}}$ and $f_{\textrm{res}}$ become comparable and a beating between them or their harmonics occurs producing a number of secondary resonances inside the 3/1 MMR.

Finally, several narrow black strips on both sides of the whole 3/1 MMR are associated to high-order mean-motion resonances, that always accompany main low-order commensurabilities at high eccentricities.


\section{Conclusions}\label{conclus}

In this paper we present a detailed analysis of the structure of the planetary 3/1 mean-motion resonance, when the outer planet is more massive than the inner one. In this case, all resonant motions are associated to a libration of the critical angle $2\sigma_1 = 3 \lambda_2 - \lambda_1 - 2\varpi_1$, while the secular angle associated to the difference in longitudes of pericenter $\Delta \varpi$, oscillates/circulates.  The composition of these two modes results in quasi-periodic oscillations of the system around a center of the 3/1 resonant domain defined by a global stationary solution, $(\pi,\pi)$--ACR. The location of an ACR is uniquely defined by two free parameters of the problem, the total angular momentum and the scaling parameter.

Applying the semi-analytical approach we analyze the topology of the Hamiltonian which describes the 3/1 MMR and obtain the families of stable symmetric and asymmetric ACRs parametrized by the planetary mass ratio. The stable symmetric families of the ($\pi,\pi$)--type are characteristic of the low and moderate eccentricity regions of the phase space, while those of the ($0,\pi$)--type located beyond the collision curve defined by initial conditions for which the planetary orbits become crossing, allowing collisions to occur. In this paper, we analyze the planetary dynamics around the ($\pi,\pi$)--ACRs.

To investigate the dynamical structures of the phase space of the 3/1 MMR, we construct dynamical maps on the representative  ($n_1/n_2$,$e_2$)--planes of initial conditions, for  low, moderate and high values of the inner planet's eccentricity, corresponding to three different ACRs, all associated to the HD60532 system. We show that the phase portrait of the 3/1 MMR is populated by a complex structure of stable modes of motion. Each ACR is shown to be the intersection of two distinct stable modes of motion, one resonant (associated to the resonant angle $2\sigma_1$) and one secular (associated to the secular angle $\Delta\varpi$). For low values of the eccentricities, secondary resonances can also be found, where the librational and the secular frequencies are commensurable. These secondary resonances in turn generate thin chaotic layers, and form a complex picture between the ACR and the edge of the resonant domain.

In addition, we detect the region of the stable quasi-resonant behaviour of the planets,  where both critical angles of the 3/1 MMR circulate, while the secular angle oscillates around zero. This region is observed for all values of the total angular momentum analyzed. This regime of the motion is topologically distinct from the pure secular regime due to opposite direction of the $\Delta\varpi$--oscillation. Therefore, the domains of unstable motion can be found separating the two regions.

\section*{Acknowledgments}
This work has been supported by the Brazilian National Research Council - CNPq (grant 153713/2010-0). The authors are grateful to Prof. Dr. S. Ferraz-Mello, Dr. J. Correa-Otto, Dr. E. Andrade-Ines, and Dr. C. Beaug\'e, for numerous suggestions and corrections to this paper. This work has made use of the facilities of the Computation Center of the University of S\~ao Paulo (LCCA-USP) and of the Laboratory of Astroinformatics (IAG/USP, NAT/Unicsul), whose purchase was made possible by the Brazilian agency FAPESP (grant 2009/54006-4) and the INCT-A.




\newpage
\section*{Appendix}

The explicit expressions for the coefficients  of the analytical first-order expansion of the Hamiltonian (\ref{hamiltoniano_laplace31}) of the 3/1 MMR:

\begin{eqnarray}
 A & = &- \dfrac{M_1}{2 J_1^3}+\dfrac{3}{2}\dfrac{M_2}{2 J_2^3}, \qquad \qquad B= - \dfrac{3}{8}\left(\dfrac{1}{2}\right)^2\dfrac{M_1}{J_1^4}, \nonumber\\
C &= &-c_2 \dfrac{M_3}{J_1 J_2^2}+\dfrac{3}{2}c_1 \dfrac {M_1}{J_2^3}, \qquad  D =- c_3 \dfrac{M_3}{J_2^3}+\dfrac{3}{2}c_1 \dfrac{M_3}{J_2^3},\nonumber\\
E & = &-c_4\dfrac{M_3}{J_2^2 \sqrt{J_1 J_2}}, \qquad \qquad \: F_1 = -\dfrac{1}{2}c_5\dfrac{M_3}{J_2^2 J_1},\label{coef}\\
F_2 & = &-\dfrac{1}{2}c_6\dfrac{M_3}{J_2^3},\qquad \qquad  \qquad F_3 = \dfrac{1}{2}c_7\dfrac{M_3}{J_2^2\sqrt{J_1 J_2}},\nonumber
\end{eqnarray}
where $M_1 = \mu_1^2 m_1^{\prime3}$, $M_2 = \mu_2^2 m_2^{\prime^3} $ and $ M_3 = m_1 M_2/m_0$, while $c_i$ are the Laplace coefficients:

\begin{eqnarray}
c_0 & = &-\dfrac{1}{2}\alpha b_{\frac{3}{2}}^{\left(1\right)}\left(\alpha\right),\nonumber\\
c_1 & = &\dfrac{1}{2}b_{\frac{1}{2}}^{\left(0\right)}\left(\alpha\right),\nonumber \\
c_2 & = & \dfrac{1}{8}\left[2\alpha D_\alpha+\alpha^2 D_\alpha^2\right]b_{\frac{1}{2}}^{\left(0\right)},\\
c_4 & = &\dfrac{1}{4}\left[2-2\alpha D_\alpha-\alpha^2 D_\alpha^2\right]b_{\frac{1}{2}}^{\left(1\right)},\nonumber
\end{eqnarray}
\begin{eqnarray}
c_5 & = & \dfrac{1}{8}\left[21+10\alpha D_\alpha+\alpha^2 D_\alpha^2\right]b_{1/2}^{\left(3\right)},\nonumber \\
c_6  & = & \dfrac{1}{4}\left[-20-10\alpha D_\alpha-\alpha^2 D_\alpha^2\right]b_{1/2}^{\left(2\right)},\nonumber\\
c_7 &= & \dfrac{1}{4}\left[17+10\alpha D_\alpha +\alpha^2 D_\alpha^2\right]b_{1/2}^{\left(1\right)}-\dfrac{27}{8}\alpha,\nonumber
\end{eqnarray}
where $D_\alpha^n$ are n-th order derivative in  $\alpha=a_1/a_2$. The coefficients $b_i$ are obtained by the series:
\begin{eqnarray}
\dfrac{1}{2}b_s^{\left(j\right)}\left(\alpha\right) & = &\dfrac{s\left(s+1\right)...\left(s+j-1\right)}{1.2.3...j}\alpha^j\label{coef_laplace} \\
& \times & \left[1+\dfrac{s\left(s+j\right)}{1\left(j+1\right)}\alpha^2+\dfrac{s\left(s+1\right)\left(s+j\right)\left(s+j+1\right)}
{1.2\left(j+1\right)\left(j+2\right)}\alpha^4+
...\right].\nonumber
\end{eqnarray}
when $j=0$,
\begin{equation}
 \dfrac{s\left(s+1\right)...\left(s+j-1\right)}{1.2.3...j}\alpha^j= 1,
\end{equation}
and the series is convergent for $\alpha<1$.

\label{lastpage}

\end{document}